\documentclass[11pt]{article}
\pdfoutput=1

\usepackage{jheppub}
\usepackage[latin1]{inputenc}
\usepackage{amsmath}
\usepackage{amsfonts}
\usepackage{amssymb,MnSymbol,slashed}
\usepackage{amsxtra}
\usepackage{color}
\usepackage{hyperref}
\hypersetup{linktocpage=true}
\usepackage{graphicx}
\usepackage{placeins}
\usepackage{upgreek} 
\usepackage{mathrsfs}
\usepackage{accents}
\usepackage{multirow} 
\usepackage{tikz}
\usepackage{calc}
\usepackage{multicol}

\usepackage{color}
\definecolor{darkgreen}{rgb}{0,0.5,0}
\definecolor{darkblue}{rgb}{0,0,0.7}
\definecolor{darkred}{rgb}{0.5,0,0.0}
\definecolor{darkorange}{rgb}{0.8,0.4,0.0}

\newcommand{\beq}{\begin{equation}}
\newcommand{\eeq}{\end{equation}}
\def\be {\begin{equation}}
\def\ee {\end{equation}}
\def\bs#1\es{\begin{split}#1\end{split}}
\def\ba#1\ea{\begin{align}#1\end{align}}
\def\baed#1\eaed{\begin{aligned}#1\end{aligned}}
\def\bged#1\eged{\begin{gathered}#1\end{gathered}}
\def\bea{\begin{eqnarray}}
\def\eea{\end{eqnarray}}

\def\nn{\nonumber}

\let\foo\bar 
\renewcommand{\bar}[1]{ {\foo{  #1} }{} }

\newlength{\dhatheight}

\newcommand{\Heckea}{H_{\Gamma_2^-}}
\newcommand{\Heckeb}{H_{\Gamma_2^+}}
\newcommand{\Heckec}{H_{\Gamma_2^0}}
\newcommand{\Ga}{\Gamma_2^-}
\newcommand{\Gb}{\Gamma_2^+}
\newcommand{\Gc}{\Gamma_2^0}

\newcommand{\T}{T=B_{89}+i V_{T^2}}
\newcommand{\U}{U=U_1 + i U_2}

\newcommand{\leemu}{ln \left(  U_2|\eta(U)|^2\right) }
\newcommand{\leemt}{ln \left(  T_2|\eta(T)|^2\right) }

\newcommand{\sz}{SL(2,\mathbb{Z})}

\newcommand{\ett}{E_8 \times E_8}

\newcommand{\et}{\hat{E}_2}
\newcommand{\sx}{SO(16) \times SO(16)}
\newcommand{\stt}{SO(8)^4}

\title{One loop amplitude for Heterotic string on $T^2$}

\author[a]{Soumya Sasmal}
\preprint{IPhT-T16/157} 

\affiliation[a]{Institut de Physique Th{\'e}orique, Universit{\'e} Paris Saclay, CEA, CNRS,\\ Orme des Merisiers, F-91191 Gif-sur-Yvette, France}

\emailAdd{soumya.sasmal@u-psud.fr, soumya.sasmal@cea.fr}

\abstract{We revisit the results of one loop string amplitude calculations for the Heterotic string theory compactified on a torus with or without Wilson lines. We give the complete elliptic genus and the harmonic part of the CP-even amplitude for the gauge groups $SO(32)$, $\ett$, $\sx$ and $SO(8)^4$.}
\begin{document}

\maketitle
%======================================================================
%\newpage
\section{Introduction}\label{sec:intro}
The one loop string amplitude calculations with half-maximal supersymmetry provide very important correction terms in the effective action of the corresponding supergravity theory. Such one loop string amplitudes in case of heterotic string theories compactified on a torus with or without Wilson lines are of profound interest in view of its duality with F-theory compactified on a K3 surface. These one loop correction terms receive no further renormalization because they serve the purpose of the anomaly cancelling term as discussed in \cite{Minasian:2016hoh}. Thus their structure may be used to extract useful informations about non-trivial axio-dilaton coupling of 7-branes in the context of the duality between heterotic on $T^2$ and F-theory on K3. The one loop string amplitude results have been calculated in parts and to serve very specific purpose in the string theory literature. In this paper, we recollect the existing results and complete  such calculations  for the cases with gauge groups $SO(32)$, $\ett$, $\sx$ and $SO(8)^4$. We provide the complete elliptic genus for these theories and present the harmonic part of these amplitudes which will prove to be of immense importance for the case of discrete $\sz$ anomaly cancellation in D=8, N=1 supergravity and put stringent consistency condition upon further compactification down to six dimensional supergravity theories with $N=(1,0)$ supersymmetry which we have discussed in \cite{Minasian:2016hoh}. The present paper may serve as a collection of the calculations which have been accomplished partially in the past with newer complements to make their results complete and their purpose more general. These calculations will be used to pave the way for the calculation of two or higher string loop calculations with half-maximal supercharges.
%=====================================================================
\section{Generalities}\label{sec:gen}
 We start our discussion by noting the field content of the Heterotic theory with gauge group $G=SO(32)$ or $\ett$ compactified on a torus $T^2$ without Wilson lines comprises of \cite{Salam:1985ns} a gravity multiplet with 1 graviton, 1 anti-symmetric two-form, 2 graviphotons, 1 real scalar ,1 gravitino, 1 dilatino and a vector multiplet in the adjoint representation of the gauge group $G=SO(32)$ or $\ett$. If in addition, we switch on non-zero Wilson lines along the cycles of the compact torus, then the gauge group $G=SO(32)$ or $\ett$ is broken down to some other gauge group like $\sx$ or $\stt$ keeping however the original rank fixed that is 16. We denote by $T$ and $U$ respectively the K\"ahler and complex structure of the torus $T^2$ such that
 
 \begin{align}
 T=T_1 +iT_2 = B_{89}+iV_{T^2},\qquad  U=U_1+iU_2.
 \end{align}
 In this article, we shall be interested in calculating the one loop string amplitude result involving 1 of the $T$ and $U$ moduli and either 4 gravitons or 4 gauge bosons or 2 gauge bosons and 2 gravitons. The CP-even amplitude follows from \cite{Ellis:1987dc,Abe:1988cq, Lerche:1987qk,Dixon:1990pc, Harvey:1995fq,Kiritsis:1997hf}
  \begin{equation}
  \mathcal{A}=t_8 V_8\int_{\mathcal{F}}\frac{d^2\tau}{\tau^2_2}\Gamma_{2,2}A(q,R,F)|_{8-forms},
  \end{equation}
  where $\Gamma_{2,2}$ is the $T^2$ lattice sum
  \begin{equation}
\Gamma_{2,2}=\frac{T_2}{\tau_2}\sum_{B \in ML(2,\mathbb{Z})}exp\left[2\pi i T det(B)-\frac{\pi T_2}{\tau_2 U_2}\vert (1 ~~U)B \begin{pmatrix}\tau\\1\end{pmatrix}\vert^2 \right]=\sum_{\vec{m},\vec{n}\in \mathbb{Z}^N} q^{P^2_L/2} \bar{q}^{P^2_R/2}.
\end{equation} 
and $A(q,R,F)|_{8-forms}$ is the elliptic genus \cite{Lerche:1987qk}. We emphasize from the very beginning that amongst many a different elegant methods available for such a calculation of amplitude, we shall be chiefly using the method of elliptic genus \cite{Lerche:1987qk, Lerche:1988np} and the method of decomposition of the  $\Gamma_{2,2}$ lattice sum into three orbits of $PSL(2,\mathbb{Z})$ \cite{Dixon:1990pc, Harvey:1995fq,Kiritsis:1997hf}. The elliptic genus $A(q,R,F)|_{8-forms}$ is an eight-form polynomial in $trR^4$, $(trR^2)^2$, $trF^4$, $(trF^2)^2$ and $trR^2trF^2$ where the lower-case ``tr" denotes the group trace in fundamental or vector representation and with coefficients some modular functions $\Phi(\tau)$ of the complex structure $\tau$ of the world-sheet torus. Thus the generic form of the elliptic genus is the following
\begin{equation}
A(q,R,F)|_{8-forms}=\Phi_1 (\tau) \frac{trR^4}{(2\pi)^4}+\Phi_2 (\tau) \frac{(trR^2)^2}{(2\pi)^4}+\Phi_3 (\tau) \frac{trF^4}{(2\pi)^4}+\Phi_4 (\tau) \frac{(trF^2)^2}{(2\pi)^4}+\Phi_5 (\tau) \frac{trR^2trF^2}{(2\pi)^4}
\end{equation} where as stated before, $\Phi_i$, $i=1,\cdots, 5$ are appropriate modular functions.
 The elliptic genus can be conveniently seen to appear from the ``gauging" of the lattice of the gauge group G. We shall next $q-$expand the modular functions $\Phi_i$ by 
 \begin{equation}
 \label{qexp}
\Phi_i({q})=\sum_{n=-1}^{\infty} c_n^i {q}^n,
\end{equation}
and decompose the $2\times 2$ matrices $B$ in the lattice sum into the  orbits of $PSL(2,\mathbb{Z})$  \cite{Dixon:1990pc, Harvey:1995fq, Kiritsis:1997hf} : 
\begin{center}

  \(\begin{array}{|c|c|c|}
\hline
 \text{Orbits} & \text{Defining properties} & \text{Canonical representative} \\
\hline
 \text{Trivial} & B=0 & \left( \begin{smallmatrix} 0&0\\ 0&0 \end{smallmatrix} \right)\\
\hline
 \text{Degenerate} & B\neq 0;~ detB=0 & \left( \begin{smallmatrix} 0&j\\ 0&p \end{smallmatrix} \right);~ j,p \neq 0.\\
\hline
\text{Non-degenerate} & B\neq 0;~ detB \neq 0 & \left( \begin{smallmatrix} k&j\\ 0&p \end{smallmatrix} \right);~ 0 \leq j < k;~ p \neq 0.\\
\hline
\end{array}\)
\end{center}
 We note however that depending on the gauge group, one may need to use subsectors of the above orbits \cite{Gutperle:1999dx, Gutperle:1999xu}. The modular integral will now look like 
 \begin{eqnarray}
  \label{int}
  \mathcal{A}&=&V_8 T_2 t_8\times \lbrace \underbrace{\int_{\mathcal{F}} \frac{d^2 \tau}{\tau_2^2}A(q,R,F)}_{\text{trivial orbit}} \\\nonumber
&{   }&  + \underbrace{\int_{\text{strip-boundary of }PSL(2,\mathbb{Z})}  \frac{d^2 \tau}{\tau_2^2} \sum_{(j,p)\neq (0,0)}e^{-\frac{\pi T_2}{\tau_2 U_2}|j+pU|^2}A(q,R,F) }_{\text{degenerate orbit}}\\\nonumber
 &{  }& +\underbrace{2\int_{\mathbb{C}^+} \frac{d^2 \tau}{\tau_2^2} \sum_{0 \leq j <k, p\neq 0} e^{-2\pi i T pk} e^{-\frac{\pi T_2}{\tau_2 U_2}|k \tau +j+pU|^2}A(q,R,F)}_{\text{non-degenerate orbit}} \rbrace .
  \end{eqnarray}
  In all the cases of amplitude calculation, we shall concentrate mostly on the harmonic part of the CP-even amplitude which is found by taking into account only the constant part of the expansion \eqref{qexp} that is the coefficient $c_0$ for the different modular functions (which we have summarized in \eqref{const} in appendix \ref{sec: modfunc}). We shall also note that the sum of the coefficients $c_{-1}$ of $q^{-1}$ vanishes in all cases so that there are no poles in the calculation. There is however the appearance of the infra-red divergence in these calculations for which we shall take appropriate renormlization scheme though we shall not detail the process here (the technical details can be found in \cite{Dixon:1990pc,Harvey:1995fq, Kiritsis:1997hf}) but just mention the result coming from it. We are keen to highlight only the non-volume suppressed harmonic part of the amplitude as the CP-odd partner of such a term in the effective action would correspond to the counter-term for the $\sz$ anomaly in the corresponding supergravity theories in 8-dimensions \cite{Minasian:2016hoh}. The non-harmonic, volume suppressed terms are world-sheet instanton corrections which can also be interpreted in the dual type I side as D-instanton corrections \cite{Kiritsis:1997hf, Kiritsis:1999ss}.

%=======================================================================
\section{Outline of the string amplitude calculation for D=8, N=1 and $G= SO(32)$ and $\ett$}\label{sec:32amp}

 In the following we shall briefly outline the principle of the calculation following the lines of \cite{Kiritsis:1997hf} so that it will be easier to understand the calculation for the cases $SO(16)^2$ and $SO(8)^4$. In all cases we shall  evaluate the CP-even amplitude which contains the curvature tensor contraction of the form $t_8 trR^4$. One can then easily derive the CP-odd sector containing the curvature contraction of the form $\epsilon_8 trR^4 = trR^{\wedge 4}$ and the harmonic part of the modular functions of T and U appearing in the CP-even sector following the method of \cite{Antoniadis:1993ze,Gregori:1997hi}.\\
  The CP-even amplitude follows from
  \begin{equation}
  \mathcal{A}=V_8t_8\int_{\mathcal{F}}\frac{d^2\tau}{\tau^2_2}\Gamma_{2,2}A(q,R,F)|_{8-forms},
  \end{equation}
  where $\Gamma_{2,2}$ is the $T^2$ lattice sum
  \begin{equation}
\Gamma_{2,2}=\frac{T_2}{\tau_2}\sum_{B \in ML(2,\mathbb{Z})}exp\left[2\pi i T det(B)-\frac{\pi T_2}{\tau_2 U_2}\vert (1 ~~U)B \begin{pmatrix}\tau\\1\end{pmatrix}\vert^2 \right]=\sum_{\vec{m},\vec{n}\in \mathbb{Z}^N} q^{P^2_L/2} \bar{q}^{P^2_R/2}.
\end{equation} 
and $A(q,R,F)|_{8-forms}$ is the elliptic genus \cite{Lerche:1987qk}
\small
\begin{subequations}
\begin{align}
A(q,R,F)^{SO(32)}&= \frac{E_4^3}{2^7 3^2 5 \eta^{24}}\frac{trR^4}{(2\pi)^4}+\frac{\et^2 E_4^2}{2^9 3^2 \eta^{24}}\frac{(trR^2)^2}{(2\pi)^4}\label{s32eg}\\\nonumber
&+\frac{trR^2 trF^2}{2^8 3^2(2\pi)^4}\left( \frac{\et E_4 E_6}{\eta^{24}}-\frac{\et^2 E_4^2}{\eta^{24}}\right)\\\nonumber
&+\frac{trF^4}{(2\pi)^4}+\frac{(trF^2)^2}{2^9 3^2(2\pi)^4}\left( \frac{E_4^3}{\eta^{24}}-\frac{2\et E_4 E_6}{\eta^{24}}+\frac{\et^2 E_4^2}{\eta^{24}}-2^7 3^2\right),\\\nonumber
A(q,R,F)^{\ett}&= \frac{E_4^3}{2^7 3^2 5 \eta^{24}}\frac{trR^4}{(2\pi)^4}+\frac{\et^2 E_4^2}{2^9 3^2 \eta^{24}}\frac{(trR^2)^2}{(2\pi)^4}\\\nonumber
&+\frac{trR^2 (trF_1^2 + trF_2^2)}{2^8 3^2(2\pi)^4}\left( \frac{\et E_4 E_6}{\eta^{24}}-\frac{\et^2 E_4^2}{\eta^{24}}\right)\\
&+\frac{trF_1^2trF_2^2}{2^8 3^2(2\pi)^4}\left( \frac{\et^2 E_4^2}{\eta^{24}}-\frac{2\et E_4 E_6}{\eta^{24}}+\frac{E_6^2}{\eta^{24}}\right)\\\nonumber
&+\frac{(trF_2^2)^2+(trF_2^2)^2}{2^8 3^2(2\pi)^4}\left( \frac{E_4^3}{\eta^{24}}-\frac{2\et E_4 E_6}{\eta^{24}}+\frac{\et^2 E_4^2}{\eta^{24}}\right).\label{e8eg}
\end{align}
\end{subequations}
\normalsize
In the above, all group traces ``tr" are in fundamental or vector representation.\\
In view of the above elliptic genus, the amplitude $\mathcal{A}$ can be viewed as the sum of integrals of the type 
\begin{equation*}
I(T,U)=\int_{\mathcal{F}}\frac{d^2\tau}{\tau_2^2}\Gamma_{2,2}(T,U)\Phi(q)
\end{equation*}
with $\Phi(q)$ being the modular form coefficient of each of the 8-form components $trR^4$, $(trR^2)^2$, $trR^2 trF^2$, $trF^4$ and $(trF^2)^2$ . \\
Next one uses the q-expansion (with $q=e^{2\pi i \tau}$) of $\Phi(q)$
\begin{equation}
\Phi({q})=\sum_{n=-1}^{\infty} c_n {q}^n,
\end{equation}
and decomposes the $2\times 2$ matrices $B$ in the lattice sum into the  orbits of $PSL(2,\mathbb{Z})$ (see \cite{Kiritsis:1997hf, Dixon:1990pc, Harvey:1995fq}) : 
\begin{center}

  \(\begin{array}{|c|c|c|}
\hline
 \text{Orbits} & \text{Defining properties} & \text{Canonical representative} \\
\hline
 \text{Trivial} & B=0 & \left( \begin{smallmatrix} 0&0\\ 0&0 \end{smallmatrix} \right)\\
\hline
 \text{Degenerate} & B\neq 0;~ detB=0 & \left( \begin{smallmatrix} 0&j\\ 0&p \end{smallmatrix} \right);~ j,p \neq 0.\\
\hline
\text{Non-degenerate} & B\neq 0;~ detB \neq 0 & \left( \begin{smallmatrix} k&j\\ 0&p \end{smallmatrix} \right);~ 0 \leq j < k;~ p \neq 0.\\
\hline
\end{array}\)
\end{center}
  The modular integration will now look like
  \begin{eqnarray}
  \label{int}
  \mathcal{A}&=&V_8 T_2 t_8\times \lbrace \underbrace{\int_{\mathcal{F}} \frac{d^2 \tau}{\tau_2^2}A(q,R,F)}_{\text{trivial orbit}} \\\nonumber
&{   }&  + \underbrace{\int_{\text{strip-boundary of }PSL(2,\mathbb{Z})}  \frac{d^2 \tau}{\tau_2^2} \sum_{(j,p)\neq (0,0)}e^{-\frac{\pi T_2}{\tau_2 U_2}|j+pU|^2}A(q,R,F) }_{\text{degenerate orbit}}\\\nonumber
 &{  }& +\underbrace{2\int_{\mathbb{C}^+} \frac{d^2 \tau}{\tau_2^2} \sum_{0 \leq j <k, p\neq 0} e^{-2\pi i T pk} e^{-\frac{\pi T_2}{\tau_2 U_2}|k \tau +j+pU|^2}A(q,R,F)}_{\text{non-degenerate orbit}} \rbrace .
  \end{eqnarray}
  To determine the leading  part (non-volume suppressed part) of the amplitude coming from the degenerate orbit, that is to evaluate the integral 
  \beq
  \int_{\text{strip-boundary of }PSL(2,\mathbb{Z})}  \frac{d^2 \tau}{\tau_2^2} \sum_{(j,p)\neq (0,0)}e^{-\frac{\pi T_2}{\tau_2 U_2}|j+pU|^2} c_0,
  \eeq
  where $c_0$ is the coefficient of $q^0$ of the $q$ expansion of the elliptic genus $A(q,R,F)$, we use result of \cite{Kiritsis:1997hf} to obtain the following harmonic part
  \bea
  \label{degint}
  \int_{\text{strip-boundary of }PSL(2,\mathbb{Z})}  \frac{d^2 \tau}{\tau_2^2} \sum_{(j,p)\neq (0,0)}e^{-\frac{\pi T_2}{\tau_2 U_2}|j+pU|^2} c_0 \\\nn
  = \left[log U_2|\eta(U)|^2 +\frac{\pi U_2}{6}\right]c_0 + \mbox{terms with $V_{T^2}$ in denominator}
  \eea
  Note that the seemingly non-harmonic $log U_2$ piece in \eqref{degint} comes from taking the appropriate renormalization scheme against the infra-red divergence of the above amplitude calculation.\\
To determine the non-volume suppressed part of the amplitude coming from the non-degenerate orbit, we use the integral  \cite{Kiritsis:1997hf}, \cite{Gutperle:1999xu}
\bea
\label{ndegint}
  T_2 \sum_{0 \leq j <k, p\neq 0} e^{-2\pi i T pk}\int_{\mathbb{C}^+} \frac{d^2 \tau}{\tau_2^2} e^{-\frac{\pi T_2}{\tau_2 U_2}|k \tau +j+pU|^2}c_0 =\\\nn
   \sum_{j} \sum_{k>0, p>0}\frac{e^{2\pi i k p T}}{k|p|} c_0+\mbox{cc.}+\mbox{volume suppressed terms.}
\eea
  
  We them sum up the leading order non-volume suppressed terms from all the three orbits which gives us
  \begin{eqnarray}
  \label{sumint}
  I(T,U)&=&\int_{\mathcal{F}}\frac{d^2\tau}{\tau_2^2}\Gamma_{2,2}(T,U)\Phi(q)\\\nonumber
  &=&\frac{\pi T_2}{3}\left[c_0-24c_{-1} \right] +\left[log U_2|\eta(U)|^2 +\frac{\pi U_2}{6}\right]c_0+\left[logT_2|\eta(T)|^2 +\frac{\pi T_2}{6}\right]c_0\\\nonumber
  &{}&+\text{non-harmonic terms with } T_2^s \text{ in denominator with s=1, 2}.
  \end{eqnarray}
  Using the above plus the $q-$expansion of different modular functions which we have summarized in \eqref{const} in \ref{sec: modfunc} we find the CP-even amplitude for SO(32)
  \begin{align}
  \label{so32cpe}
  \mathcal{A}^{SO(32)}_{\text{CP-even}}&=V_8 T_2 N \frac{\pi}{24}t_8\underbrace{\left(trR^4+\frac{1}{4}(trR^2)^2+trR^2 trF^2 + 8trF^4 \right)}_{\text{Trivial orbit}}\\\nonumber
  & +V_8 N \frac{1}{48}\left[log U_2|\eta(U)|^2 \right]\\\nonumber
  &  \times t_8\underbrace{\left(\frac{31}{15} trR^4+\frac{19}{12}(trR^2)^2+5trR^2 trF^2+2(trF^2)^2+16trF^4 \right)}_{\text{Harmonic term from the degenerate orbit}}\\\nonumber
  &  +V_8 N \frac{1}{48}\left[log T_2|\eta(T)|^2 +\frac{\pi T_2}{6}\right]\\\nonumber
  & \times  t_8\underbrace{\left(\frac{31}{15} trR^4+\frac{19}{12}(trR^2)^2+5trR^2 trF^2+2(trF^2)^2+16trF^4 \right)}_{\text{Harmonic term from the non-degenerate orbit}}\\\nonumber
  & +\text{non-harmonic terms}.
  \end{align}
Similarly, using the elliptic genus for $\ett$ we find
\begin{align}
  \label{ettcpe}
  \mathcal{A}^{\ett}_{\text{CP-even}}&=V_8 T_2 N \frac{\pi}{24}t_8\underbrace{\left(trR^4+\frac{1}{4}(trR^2)^2+trR^2 (trF_1^2+trF_2^2)-2trF_1^2trF_2^2+2(trF_1^2)^2+2(trF_2^2)^2 \right)}_{\text{Trivial orbit}}\\\nonumber
  & +V_8 N \frac{1}{48}\left[log U_2|\eta(U)|^2 \right]\\\nonumber
  &  \times t_8\underbrace{\left(\frac{31}{15} trR^4+\frac{19}{12}(trR^2)^2+5trR^2 (trF_1^2+trF_2^2)+6((trF_1^2)^2+(trF_2^2)^2) \right)}_{\text{Harmonic term from the degenerate orbit}}\\\nonumber
  &  +V_8 N \frac{1}{48}\left[log T_2|\eta(T)|^2 +\frac{\pi T_2}{6}\right]\\\nonumber
  & \times t_8\underbrace{\left(\frac{31}{15} trR^4+\frac{19}{12}(trR^2)^2+5trR^2 (trF_1^2+trF_2^2)+6((trF_1^2)^2+(trF_2^2)^2) \right)}_{\text{Harmonic term from the non-degenerate orbit}}\\\nonumber
  & +\text{non-harmonic terms}.
  \end{align}
    In both the cases above, the contributions from the trivial orbit are in fact the $T^2$ compactifications of the CP-even partner of the 10-dimensional Green-Schwarz terms \cite{Green:1984sg}
     \begin{equation}
S_{GS}^{SO(32)}=\frac{1}{192(2\pi)^5 \alpha '}\int B_2  \left(trR^{ 4}+\frac{1}{4}(trR^2)^{2}+trR^{ 2} trF^{\wedge 2} + 8trF^{4} \right)
\end{equation}
and 
 \begin{eqnarray}
   \label{e88gs}
S_{GS}^{\ett}&=&\frac{1}{192(2\pi)^5 \alpha '}\int B_2  \\\nonumber
& &\left( trR^4+\frac{1}{4}(trR^2)^2+trR^2 (trF_1^2+trF_2^2)-2trF_1^2trF_2^2+2(trF_1^2)^2+2(trF_2^2)^2 \right).
\end{eqnarray} The CP-odd partners of the non-volume suppressed harmonic terms from the degenerate and non-degenerate orbits provide with the $\sz$ anomaly cancelling term in the corresponding supergravity theory (that is D=8, N=1 SUGRA with $G=SO(32)$ or $\ett$) which have been studied in \cite{Minasian:2016hoh}.
%======================================================================
\subsection{A lift in 9 dimensions}\label{sec:32lift}

As a digression, we study the decompactification limit of the CP-even amplitude \eqref{so32cpe} to D=9 N=1 G=SO(32) theory. Suppose that the $T^2$ in the case above (section \ref{sec:32amp}) have radii $R_1$ and $R_2$ along the two cycles and the angle between them be $\omega$. We can then write the $T^2$ metric and its volume and complex structure in terms of $R_1$, $R_2$ and $\omega$ as follows
\begin{align}
\label{tormetricin9}
  G_{ij}=\left( \begin{smallmatrix} g_{88}& g_{89} \\ g_{89}& g_{99}\end{smallmatrix} \right)=\left( \begin{smallmatrix} R_1^2 &R_1 R_2 cos \omega \\ R_1 R_2 cos \omega &R_2^2\end{smallmatrix} \right)=\frac{V}{U_2}\left( \begin{smallmatrix}1 & U_1 \\ U_1 & |U|^2 \end{smallmatrix} \right).
\end{align} 

\begin{align}
\label{8to9}
V= R_1 R_2 sin \omega ,~~~~~ U_1= \frac{R_2}{R_1} cos \omega,~~~~~~ U_2=\frac{R_2}{R_1}sin \omega.
\end{align}
We use the above to decompactify the CP-even amplitude \eqref{so32cpe} by taking $\omega = \frac{\pi}{2}$ and $R_2=V_1$ such that $V_9 = V_8 R_2$ and $V_{10}=V_8 R_1 R_2$ are the normalized world-volumes in D=9 and D=10 respectively. In this limit $U_1=0$ , $U_2 = R_2/R_1$, $log U_2|\eta(U)|^2 = -\frac{\pi U_2}{6}$ and the limit of the amplitude \eqref{so32cpe} gives
\begin{eqnarray}
  \label{so32d9}
  \mathcal{A}^{SO(32)}_{\text{CP-even}}&=&V_9 R_1 N t_8\frac{\pi}{24}\left(trR^4+\frac{1}{4}(trR^2)^2+trR^2 trF^2 + 8trF^4 \right)\\\nonumber
  &{}& +V_9 N \frac{1}{48}\frac{1}{R_1}t_8\left(\frac{31}{15} trR^4+\frac{19}{12}(trR^2)^2+5trR^2 trF^2+2(trF^2)^2+16trF^4 \right)\\\nonumber
  &{}&+\text{non-harmonic terms}.
  \end{eqnarray}
We can compare the above with the direct calculation of the string amplitude in D=9 as is calculated in \cite{Kiritsis:1999ss}
\begin{eqnarray}
\label{so32d91}
\mathcal{A}^{SO(32)}_{\text{CP-even}}&=& V_{10}\lbrace N t_8\left(trR^4+\frac{1}{4}(trR^2)^2+trR^2 trF^2 + 8trF^4 \right)\\\nonumber
&{}&+\frac{N_1}{R^2_1}t_8\left(\frac{31}{15} trR^4+\frac{19}{12}(trR^2)^2+5trR^2 trF^2+2(trF^2)^2+16trF^4 \right)\\\nonumber
  &{}&+\frac{N_2}{R_1^4}t_8\left(3(trF^2)^2+5trR^2 trF^2 +2 (trR^2)^2 \right) +\frac{N_3}{R_1^6}t_8\left(trR^2+TrF^2 \right)^2 \rbrace
\end{eqnarray} and we see that the first two lines match as they should. The polynomials in the 3rd line of \eqref{so32d91} are also present in the volume suppressed part of \eqref{so32cpe}. In both \eqref{so32d9} and \eqref{so32d91} the first line is the circle compactification of the CP-even Green-Schwarz term  and the second line contains the CP-even partner of the $\sz$ anomaly cancelling term in D=8 as we have seen. However in D=9 this term is completely harmless as there is no chiral anomaly in D=9. Nonetheless, it is interesting to see the presence of this term in D=9 amplitude result which upon further compactification on $S^1$ shall give rise to the anomaly cancelling term in D=8.

%========================================================================
\section{String amplitude with G= $SO(16) \times SO(16)$}\label{sec:G16}

Now we consider D=10 Heterotic string theory with gauge group $\ett$ compactified on a $T^2$ with K\"ahler structure $\T$ and complex structure $\U$ and with the following Wilson line on $T^2$
\begin{equation}
Y_i^1 = (0^4 , \frac{1}{2}^4, 0^4 , \frac{1}{2}^4), \quad \quad Y_i^2= (0^8,0^8),\qquad i=1,\cdots ,16,
\end{equation}
so that the gauge group is broken to $SO(16) \times SO(16)$ in D=8. One can of course rearrange the 8 non-zero values of the Wilson lines so that one can start from $SO(32)$ gauge group in D=10 and again obtain $SO(16) \times SO(16)$ in D=8.\\

We now discuss the group decomposition $\ett \supset \sx $   which we shall find extremely useful to understand the string amplitude part.\\
For the decomposition $\ett \supset \sx $ we have
\begin{equation}
\textbf{248 }\oplus \textbf{ 248}=\underbrace{\textbf{(120,1) $\oplus$ (1,120) }}_{\text{adjoint rep. of }\sx}\oplus \underbrace{\textbf{ (128,1) $\oplus$ (1,128)}}_{\text{spinor rep. of }\sx}.
\end{equation}
The rules for transcribing group trace ``Tr" in the adjoint representation towards the group trace ``tr" in the fundamental representation for $SO(N)$ groups \cite{vanNieuwenhuizen:1989gc, Erler:1993zy}
\begin{align}
   TrF^2_{SO(N)}&= (N-2) ~trF^2_{SO(N)} \label{trace1}\,,\\
    TrF^4_{SO(N)}&= (N-8)~trF^4_{SO(N)}~ +~ 3~(trF^2_{SO(N)})^2 \label{trace3}\,.
   \end{align}
For the $\textbf{ (128,1) $\oplus$ (1,128)} $ representation, we write the traces formula
\begin{subequations}
\begin{align}
tr_{128}F_1^2 + tr_{128}F_2^2 &= 16 trF_1^2 +16 trF_2^2,\label{16trace1}\\
tr_{128}F_1^4 + tr_{128}F_2^4 &= 6 (trF_1^2)^2 + 6 (trF_2^2)^2 -8trF_1^4 -8trF_2^4.\label{16trace2}
\end{align}
\end{subequations}
For the sake of completeness, we also provide the branching rule for the  decomposition $SO(32)\supset \sx$ 
\begin{equation}
\textbf{496}=\underbrace{\textbf{(120,1) $\oplus$ (1,120) }}_{\text{adjoint rep. of }\sx}\oplus \underbrace{\textbf{ (16,16)}}_{\text{cospinor rep. of }\sx}.
\end{equation}
For the $\textbf{(16,16)}$ representation, we write the traces formula
\begin{subequations}
\begin{align}
tr_{(16,16)}F^2&= 16 trF_1^2 +16 trF_2^2,\label{16trace3}\\
tr_{(16,16)}F^4 &= 16trF_1^4+16trF_2^4+6 (trF_1^2)(trF_2^2).\label{16trace4}
\end{align}
\end{subequations}

We now elaborate the process of the CP-even 5-point string amplitude for the $\sx$ following the lines of \cite{Gutperle:1999dx} where the pieces of the calculation have been provided e.g. the coefficient of $trR^4$, $trF^4$ and $(trF^2)^2$ for the non-degenerate orbit \eqref{int}. We shall provide the CP-even part of the amplitude in the leading order non-volume suppressed harmonic forms in trivial, degenerate and non-degenerate orbits. \\
The amplitude will be derived from
\begin{equation}
\label{16int}
  \mathcal{A}= V_8t_8\int_{\mathcal{F}}\frac{d^2\tau}{\tau^2_2}\Gamma_{2,2}A(q,R,F)|_{8-forms},
  \end{equation}
  where $\Gamma_{2,2}$ is the $T^2$ lattice sum as before
  \begin{equation}
\Gamma_{2,2}=\frac{T_2}{\tau_2}\sum_{B \in ML(2,\mathbb{Z})}exp\left[2\pi i T det(B)-\frac{\pi T_2}{\tau_2 U_2}\vert (1 ~~U)B \begin{pmatrix}\tau\\1\end{pmatrix}\vert^2 \right]=\sum_{\vec{m},\vec{n}\in \mathbb{Z}^N} q^{P^2_L/2} \bar{q}^{P^2_R/2}
\end{equation} 
with $B$ being the $2 \times 2$ matrix
\begin{equation}
B=\left(\begin{matrix}
m_1 & \quad n_1\\
m_2 & \quad n_2
\end{matrix} \right). 
\end{equation}
The form of the elliptic genus $A(q,R,F)$ shall depend on the spin structure as we shall describe shortly and hence there are 3 different elliptic genus for trivial, degenerate and non-degenerate orbits which we shall note by $A_{\text{trivial}}(q,R,F)$, $A_{\text{degenerate}}(q,R,F)$ and $A_{\text{non-degenerate}}(q,R,F)$ respectively. The general elliptic genus is obtained from the gauging of \cite{Lerche:1987qk} 
\begin{equation}
\label{16eg}
A(q,R,F)=\frac{1}{\eta^{24}}Exp\left(\frac{trR^2}{(2\pi)^2}\frac{\et}{48} \right) \times Exp\left(\frac{trR^4}{(2\pi)^4}\frac{E_4}{2^7 3^2 5} \right)\times \sum_{a,b=1}^2 \underbrace{ \theta^8\left[ \substack{a \\ b}\right]}_{SO(16)_1} \times \underbrace{\theta^8\left[ \substack{a+m_1 \\ b+n_1}\right]}_{SO(16)_2}.
\end{equation}
We have summarised our convention of Jacobi theta functions in appendix \ref{sec: modfunc}.  The labels $SO(16)_1$ and $SO(16)_2$ in \eqref{16eg} denote the  gauging of the theta functions according to two $SO(16)$s.\\
The trivial orbit is characterised by $B=0$ so that the elliptic genus will be 
\begin{eqnarray}
& &A_{\text{trivial}}(q,R,F)=\\\nn
& &\frac{E_4^3}{2^7 3^2 5 \eta^{24}}\frac{trR^4}{(2\pi)^4}+\frac{\et^2 E_4^2}{2^9 3^2 \eta^{24}}\frac{(trR^2)^2}{(2\pi)^4}\\\nn
& &+\frac{trR^2 (trF_1^2 + trF_2^2)}{2^8 3^2(2\pi)^4}\left( \frac{\et E_4 E_6}{\eta^{24}}-\frac{\et^2 E_4^2}{\eta^{24}}\right)\\\nonumber
& &+\frac{trF_1^2trF_2^2}{2^8 3^2(2\pi)^4}\left( \frac{\et^2 E_4^2}{\eta^{24}}-\frac{2\et E_4 E_6}{\eta^{24}}+\frac{E_6^2}{\eta^{24}}\right)\\\nonumber
& &+\frac{(trF_1^2)^2+(trF_2^2)^2}{2^8 3^2(2\pi)^4}\left( \frac{E_4^3}{\eta^{24}}-\frac{2\et E_4 E_6}{\eta^{24}}+\frac{\et^2 E_4^2}{\eta^{24}}\right).
\end{eqnarray}
The degenerate orbit is characterised by $B \neq 0,\quad det(B)=0$, for which choose the two following sectors 
\begin{equation}
\label{16degsect}
B^{(1)}=\left(\begin{matrix} 0 & ~~~~2j\\ 0 & ~~p \end{matrix} \right),\quad \quad B^{(2)}=\left(\begin{matrix} 0 & ~~~~2j+1\\ 0 & ~~p \end{matrix} \right),\quad j, p\in \mathbb{Z}.
\end{equation}
 For the gauging, we use the identities \eqref{thetapartition} and the definitions of Eisenstein series  given in \eqref{eisenstein} plus the combinations $f_1$, $f_2$, $f_3$ of theta functions 

\begin{align}
\label{deff1}
f_1 =\theta_3^4 + \theta_4^4,\quad f_2 =\theta_2^4 - \theta_4^4,\quad f_3 =-\theta_2^4-\theta_3^4.
\end{align}
  
  The elliptic genus for degenerate orbit is then
%\small
\begin{eqnarray}
\label{16egdeg}
& & A_{\text{degenerate}}(q,R,F) =\\\nonumber
& &\frac{trR^4}{(2\pi)^4} \frac{E_4}{2^7 3^2 5 \eta^{24}}\left( B^{(1)} \sum_{a=2}^4 \theta^{16}_a +B^{(2)}2\theta_3^8 \theta_4^8\right)\\\nn
&& + \frac{(trR^2)^2}{(2\pi)^4} \frac{\et^2}{2^9 3^2  \eta^{24}}\left( B^{(1)} \sum_{a=2}^4 \theta^{16}_a +B^{(2)}2\theta_3^8 \theta_4^8\right)\\\nonumber
& &- \frac{trR^2(trF_1^2 +trF_2^2)}{2^8 3^2(2\pi)^4  \eta^{24}} \lbrace 2B^{(1)} (\et E_4 E_6-\et^2 E_4^2)-B^{(2)}(\et\theta_3^8 \theta_4^8 )(f_2 + f_3+2\et)\rbrace\\\nonumber
& &+\frac{trF_1^4 +trF_2^4}{2^7 3 (2\pi)^4 \eta^{24}} \lbrace B^{(1)} (-\theta_2^{16}\theta_3^4 \theta_4^4 +\theta_3^{16}\theta_2^4 \theta_4^4 -\theta_4^{16}\theta_2^4 \theta_3^4)+ B^{(2)}(\theta_3^8 \theta_4^8(\theta_2^4 \theta_4^4-\theta_2^4 \theta_3^4))\rbrace\\\nonumber
& &+ \frac{(trF_1^2)^2 +(trF_2^2)^2}{2^9 3^2 (2\pi)^4 \eta^{24}}\lbrace B^{(1)}\sum_{a=2}^4 \theta^{16}_a (\et + f_{a-1})^2 +B^{(2)}\theta_3^8 \theta_4^8 \left[(f_2 +\et)^2 +(f_3+\et)^2 \right]  \rbrace \\\nonumber
& &+\frac{(trF_1^2)(trF_2^2)}{2^8 3^2 (2\pi)^4 \eta^{24}}\lbrace B^{(1)}\sum_{a=2}^4 \theta^{16}_a (\et+f_{a-1})^2 +B^{(2)}\theta_3^8 \theta_4^8 \left[(f_2 +\et)^2 +(f_3+\et)^2 -9\theta_2^8\right]  \rbrace.
\end{eqnarray}
%\normalsize
In the above elliptic genus the $ B^{(1)}$ and $ B^{(2)}$ act as operators such that in the amplitude integration \eqref{16int} one should take into account the values of the matrix $B$ as given in \eqref{16degsect}.\\
Finally we come to the non-degenerate orbit ($B\neq 0$, $det(B)\neq 0$) whose matrix representative is 
\begin{equation*}
B=\left( \begin{smallmatrix} k&j\\ 0&p \end{smallmatrix} \right);\qquad 0 \leq j < k;~ p \neq 0.
\end{equation*} We have to use the following 4 sectors of this representative matrix because of the spin structure \eqref{16eg}
\begin{eqnarray}
\label{16ndegsect}
B^{(1)}&=&\left(\begin{matrix} 2k & ~~~~2j\\ 0 & ~~p \end{matrix} \right),\quad \quad B^{(2)}=\left(\begin{matrix} 2k & ~~~~2j+1\\ 0 & ~~p \end{matrix} \right),\\\nonumber
B^{(3)}&=&\left(\begin{matrix} 2k+1 & ~~~~2j\\ 0 & ~~p \end{matrix} \right),\quad \quad B^{(4)}=\left(\begin{matrix} 2k+1 & ~~~~2j+1\\ 0 & ~~p \end{matrix} \right),\qquad 0 \leq j < k,\quad j,k,p\in \mathbb{Z}.
\end{eqnarray}
The elliptic genus for the non-degenerate orbit is then
\small
\begin{eqnarray}
\label{16egndeg}
&& A_{\text{non-degenerate}}(q,R,F) =\\\nonumber
&& \frac{trR^4}{(2\pi)^4} \frac{E_4}{2^7 3^2 5 \eta^{24}}\lbrace  B^{(1)} \sum_{a=2}^4 \theta^{16}_a +B^{(2)}2\theta_3^8 \theta_4^8 + B^{(3)}2\theta_2^8 \theta_3^8 +B^{(4)}2\theta_2^8 \theta_4^8 \rbrace \\\nonumber
&&\\\nn
&&  +\frac{(trR^2)^2}{(2\pi)^4} \frac{\et^2}{2^9 3^2  \eta^{24}}\lbrace B^{(1)} \sum_{a=2}^4 \theta^{16}_a +B^{(2)}2\theta_3^8 \theta_4^8 + B^{(3)}2\theta_2^8 \theta_3^8 +B^{(4)}2\theta_2^8 \theta_4^8 \rbrace\\\nonumber
&&\\\nn
&&- \frac{trR^2(trF_1^2 +trF_2^2)}{2^8 3^2(2\pi)^4  \eta^{24}} \lbrace(  2B^{(1)} (\et E_4 E_6-\et^2 E_4^2) -B^{(2)}(\et\theta_3^8 \theta_4^8 )(f_2 + f_3+2\et)\\\nonumber
&&\text{\qquad \qquad \qquad \qquad \qquad}-B^{(3)}(\et\theta_2^8 \theta_3^8 )(f_1 + f_2+2\et)-B^{(4)}(\et\theta_2^8 \theta_4^8 )(f_1 + f_3+2\et)\rbrace \\\nonumber
&&\\\nn
&&+\frac{trF_1^4 +trF_2^4}{2^7 3 (2\pi)^4 \eta^{24}} \lbrace B^{(1)} (-\theta_2^{16}\theta_3^4 \theta_4^4 +\theta_3^{16}\theta_2^4 \theta_4^4 -\theta_4^{16}\theta_2^4 \theta_3^4) + B^{(2)}(\theta_3^8 \theta_4^8(\theta_2^4 \theta_4^4-\theta_2^4 \theta_3^4))\\\nonumber
&&\text{\qquad \qquad \qquad \qquad \qquad}+ B^{(3)}(\theta_2^8 \theta_3^8(\theta_2^4 \theta_4^4-\theta_3^4 \theta_4^4))+ B^{(4)}(\theta_2^8 \theta_4^8(-\theta_3^4 \theta_4^4-\theta_2^4 \theta_3^4))\rbrace\\\nonumber
&&\\\nn
&&+ \frac{(trF_1^2)^2 +(trF_2^2)^2}{2^9 3^2 (2\pi)^4 \eta^{24}}\lbrace B^{(1)}\sum_{a=2}^4 \theta^{16}_a (\et+f_{a-1})^2 +B^{(2)}\theta_3^8 \theta_4^8 \left[(f_2 +\et)^2 +(f_3+\et)^2 \right]\\\nonumber
&&\text{\qquad \qquad \qquad \qquad \qquad} +B^{(3)}\theta_2^8 \theta_3^8 \left[(f_1 +\et)^2 +(f_2+\et)^2 \right] +B^{(4)}\theta_2^8 \theta_4^8 \left[(f_1 +\et)^2 +(f_3+\et)^2 \right]\rbrace \\\nonumber
&&\\\nn
&&+\frac{(trF_1^2)(trF_2^2)}{2^8 3^2 (2\pi)^4 \eta^{24}}\lbrace B^{(1)}\sum_{a=2}^4 \theta^{16}_a (\et+f_{a-1})^2 +B^{(2)}\theta_3^8 \theta_4^8 \left[(f_2 +\et)^2 +(f_3+\et)^2 -9\theta_2^8\right]\\\nonumber
&&\text{\qquad \qquad \qquad} +B^{(3)}\theta_2^8 \theta_3^8 \left[(f_1 +\et)^2 +(f_2+\et)^2 -9\theta_4^8\right] +B^{(4)}\theta_2^8 \theta_4^8 \left[(f_1 +\et)^2 +(f_3+\et)^2 -9\theta_3^8\right]\rbrace.
\end{eqnarray}
\normalsize
Once again in the above, the terms $B^{(i)}$ with $i=1,2,3,4$ denote the sector operators so that one takes into account correctly the values of the matrix elements $B$ according to the convention \eqref{16ndegsect}.\\
The complete amplitude is then
\begin{eqnarray}
  \label{int2}
  \mathcal{A}&=& T_2 V_8 t_8\times \lbrace \int_{\mathcal{F}} \frac{d^2 \tau}{\tau_2^2}A(q,R,F)_{\text{trivial}} \\\nonumber
&{   }&  + \int_{\text{strip-boundary of }PSL(2,\mathbb{Z})}  \frac{d^2 \tau}{\tau_2^2} \sum_{(n_1,n_2)\neq (0,0)}e^{-\frac{\pi T_2}{\tau_2 U_2}|n_1+n_2 U|^2}A(q,R,F)_{\text{degenerate}} \\\nonumber
 &{  }& +2\int_{\mathbb{C}^+} \frac{d^2 \tau}{\tau_2^2} \sum_{0 \leq n_1 <m_1, n_2\neq 0} e^{-2\pi i T pk} e^{-\frac{\pi T_2}{\tau_2 U_2}|m_1 \tau +n_1+n_2 U|^2}A(q,R,F)_{\text{non-degenerate}} \rbrace ,\\
 &=& \mathcal{A}_{\text{trivial}}+\mathcal{A}_{\text{degenerate}}+\mathcal{A}_{\text{non-degenerate}}
  \end{eqnarray}
  The trivial orbit amplitude gives
  \begin{align}
  \label{161}
  \mathcal{A}_{\text{trivial}}&= T_2 V_8 t_8 \frac{1}{(2\pi)^4}\lbrace trR^4+\frac{1}{4}(trR^2)^2+trR^2 (trF_1^2+trF_2^2)\\\nonumber
  &{}-2trF_1^2trF_2^2+2(trF_1^2)^2+2(trF_2^2)^2\rbrace
  \end{align}
  To evaluate the degenerate amplitude, we q-expand the modular function in the elliptic genus \eqref{16egdeg} and take the constant coefficients which we have noted in \eqref{const} which shall provide the harmonic part of the amplitude. In this respect we also note that we can sum up the contributions of $B^{(1)}$ and $B^{(2)}$ sectors in \eqref{16degsect} so that the sum run in the complete set of integers for $n_1$ and $n_2$ so that the CP-even modular coefficient will be $log U_2 |\eta(U)|^2$ (the $logU_2$ follows from the renormalization scheme). Also we note that the sum of the coefficients of $1/q$ is zero so that there are no poles in $q$.  Using \eqref{degint} we find the harmonic part of the CP-even amplitude coming from the degenerate orbit
  \begin{align}
  \label{162}
  \mathcal{A}_{\text{degenerate}}= \frac{1}{(2\pi)^4} log U_2 |\eta(U)|^2 &V_8 t_8\lbrace \frac{488}{360} trR^4+\frac{200}{288}(trR^2)^2+\frac{7}{3}trR^2 (trF_1^2+trF_2^2)\\\nonumber
&   +\frac{16}{3}(trF_1^4+trF_2^4)+2((trF_1^2)^2+(trF_2^2)^2)\rbrace.
  \end{align}
 
  Finally for the non-degenerate amplitude we again q-expand the modular functions in the elliptic genus \eqref{16egndeg} and check that there is no pole in $q$. Next we note that the leading term in the harmonic part for $B^{(1)}$ and $B^{(2)}$ sectors are the same and is equal to $\left[log T_2|\eta(2T)|^2 +\frac{\pi T_2}{3}\right]$. We then sum the constant coefficients  which shall provide the leading term (which are not volume suppressed) in the harmonic part. The constant coefficients in $B^{(3)}$ and $B^{(4)}$ are the same and hence the sum over $m_1$, $n_1$ and $n_2$ can be extended to the complete $\mathbb{Z}$ with the contribution $\left[log T_2|\eta(2T)|^2 -log T_2|\eta(T)|^2+\frac{\pi T_2}{6}\right]$. Once again, we evaluate the CP-even integral using \eqref{ndegint} and the leading term (harmonic) in the non-degenerate amplitude will be (we write only the non-volume suppressed harmonic part of the amplitude)
  \begin{eqnarray}
  \label{163}
& &  \mathcal{A}_{\text{non-degenerate}}=\\\nn
&&\frac{1}{(2\pi)^4}\left[log T_2|\eta(2T)|^2 +\frac{\pi T_2}{3}\right] V_8 t_8\lbrace \frac{488}{360} trR^4+\frac{200}{288}(trR^2)^2+\frac{7}{3}trR^2 (trF_1^2+trF_2^2)\\\nonumber
&&\text{\qquad \qquad \qquad \qquad \qquad \qquad \qquad \qquad} +\frac{16}{3}(trF_1^4+trF_2^4)+2((trF_1^2)^2+(trF_2^2)^2)\rbrace\\\nonumber
&  & + \frac{1}{2(2\pi)^4}\left[log T_2|\eta(2T)|^2 -log T_2|\eta(T)|^2+\frac{\pi T_2}{6}\right]\times \\\nn
&& \text{\qquad \qquad \qquad \qquad \qquad \qquad \qquad} V_8 t_8\lbrace 256\left( \frac{trR^4}{360} + \frac{(trR^2)^2}{288}\right)+\frac{8}{3}trR^2(trF_1^2 +trF_2^2)\\\nonumber
&  &\text{\qquad \qquad \qquad \qquad \qquad \qquad \qquad \qquad}-\frac{16}{3}(trF_1^4 +trF_2^4)+4((trF_1^2)^2+(trF_2^2)^2) \rbrace
  \end{eqnarray}
We see again that the trivial orbit amplitude \eqref{161} is the compactification of the $\ett$ Green-Schwarz term \eqref{e88gs} such that $\ett$ is broken down to $\sx$. The polynomial
\begin{equation}
Y_8=\frac{488}{360} trR^4+\frac{200}{288}(trR^2)^2+\frac{7}{3}trR^2 (trF_1^2+trF_2^2) +\frac{16}{3}(trF_1^4+trF_2^4)+2((trF_1^2)^2+(trF_2^2)^2)
\end{equation}
is in fact the 8-form polynomial in the $\sz$ anomaly counter-term \cite{Minasian:2016hoh} and is due to the fermions in the D=8, N=1 SUGRA which transform under the adjoint representation in $\sx$. The other 8-form polynomial
\begin{equation}
Y_8'=256\left( \frac{trR^4}{360} + \frac{(trR^2)^2}{288}\right)+\frac{8}{3}trR^2(trF_1^2 +trF_2^2)-\frac{16}{3}(trF_1^4 +trF_2^4)+4((trF_1^2)^2+(trF_2^2)^2) 
\end{equation}
is due to the massive vector multiplet in $\textbf{ (128,1) $\oplus$ (1,128)} $ representation of $\sx$.
%========================================================================================
\section{String amplitude with G= $SO(8)^4$}\label{sec:G8}

Finally we come to the case of the D=8, N=1 theory with gauge group $SO(8)^4$ which can be obtained from D=10, N=1 theory with gauge group either $SO(32)$ or $\ett$ compactified on a $T^2$ with appropriate Wilson lines along the two 1-cycles of the torus. Before plunging into the details of the string loop calculation we first note the group traces originating from the group decompositions $SO(32) \rightarrow SO(8)^4$ and $\ett \rightarrow SO(8)^4$ which shall prove indispensable to understand the string loop amplitude.\\
For the decomposition $E_8 \supset SO(8)^2$ we have
\begin{align}
\textbf{248}=\underbrace{\textbf{(28,1) $\oplus$ (1,28) }}_{\text{adjoint rep. of }SO(8) \times SO(8)}
&\oplus\underbrace{\textbf{(8,8)}}_{\text{bifundamental rep. of }SO(8) \times SO(8)} \\\nonumber
& \oplus \underbrace{\textbf{(8,8)'}}_{\text{spinor rep. of }SO(8)\times SO(8)}\oplus \underbrace{\textbf{(8,8)''}}_{\text{cospinor rep. of }SO(8) \times SO(8)}.
\end{align} 
Thus the complete decomposition $E_8^{(1)} \rightarrow SO(8)_{(1)} \times SO(8)_{(2)}$ plus $E_8^{(2)} \rightarrow SO(8)_{(3)} \times SO(8)_{(4)}$ gives
\begin{align}
\label{e8decomp}
\textbf{248}\oplus \textbf{248} =& \textbf{(28,1,1,1)}\oplus \textbf{(1,28,1,1)}\oplus \textbf{(1,1,28,1)}\oplus \textbf{(1,1,1,28)}\\\nonumber
&\oplus \textbf{(8,8,1,1)}\oplus \textbf{(1,1,8,8)}\\\nonumber
& \oplus \textbf{(8,8,1,1)}'\oplus \textbf{(1,1,8,8)}'\\\nonumber
& \oplus \textbf{(8,8,1,1)}''\oplus \textbf{(1,1,8,8)}''.
\end{align}
For the decomposition $SO(32) \rightarrow SO(8)_{(1)} \times SO(8)_{(2)} \times SO(8)_{(3)} \times SO(8)_{(4)}$ we have
\begin{align}
\label{32decomp}
\textbf{496} =& \textbf{(28,1,1,1)}\oplus \textbf{(1,28,1,1)}\oplus \textbf{(1,1,28,1)}\oplus \textbf{(1,1,1,28)}\\\nonumber
&\oplus \textbf{(8,8,1,1)}\oplus \textbf{(1,1,8,8)}\\\nonumber
& \oplus \textbf{(8,1,8,1)}\oplus \textbf{(1,8,1,8)}\\\nonumber
& \oplus \textbf{(1,8,8,1)}\oplus \textbf{(8,1,1,8)}.
\end{align}
From the decomposition \eqref{e8decomp} we see that $E_8^{(1)} \rightarrow SO(8)_{(1)} \times SO(8)_{(2)}$ plus $E_8^{(2)} \rightarrow SO(8)_{(3)} \times SO(8)_{(4)}$ has a preferred $trF_1^2 trF_2^2$ and $trF_3^2 trF_4^2$ interaction. The T-duality exchanges the spinor and co-spinor representation with the bi-fundamental representations and we shall see that this fact appears in the string 1-loop elliptic genus as orbifold shifts \cite{Kiritsis:2000zi} which gives the mixed interaction of the type $trF_1^2 trF_3^2$ and $trF_1^2 trF_4^2$  etc, even if one starts with the decomposition $E_8^{(1)} \rightarrow SO(8)_{(1)} \times SO(8)_{(2)}$ and $E_8^{(2)} \rightarrow SO(8)_{(3)} \times SO(8)_{(4)}$. \\
We finally summarize the trace formula for different states \cite{vanNieuwenhuizen:1989gc, Erler:1993zy}
\begin{subequations}
\begin{align}
\label{8trace}
& Tr_{28}F^2 = 6 trF^2, \quad Tr_{28}F^4= 3(trF^2)^2,\\
& tr_{(8,8)}F^2= 8trF_1^2 +8trF_2^2, \quad tr_{(8,8)}F^4=8trF_1^4 +8trF_2^4+6trF_1^2trF_2^2,\\
& tr_{(8,8)'}F^2=tr_{(8,8)''}F^2= 8trF_1^2 +8trF_2^2,\\
& tr_{(8,8)'}F^4=tr_{(8,8)''}F^4 = 3(trF_1^2)^2+3(trF_2^2)^2 +6trF_1^2trF_2^2 -4trF_1^4 -4trF_2^4.
\end{align}
\end{subequations}
With the above details behind we shall now describe the calculation of the CP-even amplitude by the q-expansion method \cite{Gutperle:1999xu} as in the case of $\sx$ (\ref{sec:G16}). \\
As before, the amplitude has the generic form
\begin{equation}
\label{8int}
  \mathcal{A}= V_8 t_8\int_{\mathcal{F}}\frac{d^2\tau}{\tau^2_2}\Gamma_{2,2}A(q,R,F)|_{8-forms},
  \end{equation}
  where $\Gamma_{2,2}$ is the $T^2$ lattice sum 
  \begin{equation}
\Gamma_{2,2}=\frac{T_2}{\tau_2}\sum_{B \in ML(2,\mathbb{Z})}exp\left[2\pi i T det(B)-\frac{\pi T_2}{\tau_2 U_2}\vert (1 ~~U)B \begin{pmatrix}\tau\\1\end{pmatrix}\vert^2 \right]=\sum_{\vec{m},\vec{n}\in \mathbb{Z}^N} q^{P^2_L/2} \bar{q}^{P^2_R/2}
\end{equation} 
with $B$ being the $2 \times 2$ matrix
\begin{equation}
B=\left(\begin{matrix}
m_1 & \quad n_1\\
m_2 & \quad n_2
\end{matrix} \right). 
\end{equation}
To define the elliptic genus we shall start with the D=10 N=1 $\ett$ Heterotic string compactified on a $T^2$ with the Wilson line 
\begin{equation}
Y_i^1 = (0^4 ,\frac{1}{2}^4 , 0^4,  \frac{1}{2}^4), \quad Y_i^2= (0^4 , \frac{1}{2}^4, 0^4, \frac{1}{2}^4),\qquad i=1,\cdots ,16,
\end{equation}
so that the gauge group decomposition $E_8^{(1)} \rightarrow SO(8)_{(1)} \times SO(8)_{(2)}$ and $E_8^{(2)} \rightarrow SO(8)_{(3)} \times SO(8)_{(4)}$ applies. Thus the elliptic genus is obtained by gauging
\begin{eqnarray}
\label{8eg}
A(q,R,F)&=&\frac{1}{\eta^{24}}Exp\left(\frac{trR^2}{(2\pi)^2}\frac{\et}{48} \right) \times Exp\left(\frac{trR^4}{(2\pi)^4}\frac{E_4}{2^7 3^2 5} \right)\\\nn
&&\times \sum_{a,b=1}^2 \underbrace{ \theta^4\left[ \substack{a \\ b}\right] \theta^4\left[ \substack{a+m_2 \\ b+n_2}\right]}_{SO(8)_{(1)} \times SO(8)_{(2)}} \times \underbrace{\theta^4\left[ \substack{a+m_1 \\ b+n_1}\right]\theta^4\left[ \substack{a+m_1+m_2 \\ b+n_1+n_2}\right]}_{SO(8)_{(3)} \times SO(8)_{(4)}}.
\end{eqnarray}
In the above \eqref{8eg} we have labelled the theta functions by $SO(8)_{(1)} \times SO(8)_{(2)}$ and $ SO(8)_{(3)} \times SO(8)_{(4)}$ to denote that those functions are to be ``gauged" accordingly by the 4 copies of $SO(8)$s.
We now decompose the integration by now familiar method of the decomposition to trivial, degenerate and non-degenerate orbit. The elliptic genus for the trivial orbit ($B=0$) shall be
\small
\begin{eqnarray}
\label{8triveg}
&&A_{\text{trivial}}(q,R,F)=\\\nn
&&\frac{E_4^3}{2^7 3^2 5 \eta^{24}}\frac{trR^4}{(2\pi)^4}+\frac{\et^2 E_4^2}{2^9 3^2 \eta^{24}}\frac{(trR^2)^2}{(2\pi)^4}+\frac{trR^2 \sum_{i=1}^4 trF_i^2 }{2^8 3^2(2\pi)^4}\left( \frac{\et E_4 E_6}{\eta^{24}}-\frac{\et^2 E_4^2}{\eta^{24}}\right)\\\nonumber
&&+\frac{(trF_1^2trF_3^2+trF_2^2trF_4^2+trF_1^2trF_4^2+trF_2^2trF_3^2)}{2^8 3^2(2\pi)^4}\left( \frac{\et^2 E_4^2}{\eta^{24}}-\frac{2\et E_4 E_6}{\eta^{24}}+\frac{E_6^2}{\eta^{24}}\right)\\\nonumber
&&+\frac{\sum_{i=1}^4 (trF_i^2)^2}{2^8 3^2(2\pi)^4}\left( \frac{E_4^3}{\eta^{24}}-\frac{2\et E_4 E_6}{\eta^{24}}+\frac{\et^2 E_4^2}{\eta^{24}}\right)\\\nonumber
&&+\frac{trF_1^2trF_2^2+trF_3^2trF_4^2}{2^7 3^2(2\pi)^4}\left( \frac{E_4^3}{\eta^{24}}-\frac{2\et E_4 E_6}{\eta^{24}}+\frac{\et^2 E_4^2}{\eta^{24}}\right).
\end{eqnarray}
\normalsize
For the degenerate orbit ($B \neq 0$ and $det(B)=0$) we choose the following sectors
\small
\begin{eqnarray}
\label{8degsect}
B^{(1)}&=&\left(\begin{matrix} 0 & ~~~~2j\\ 0 & ~~2p \end{matrix} \right),\quad \quad B^{(2)}=\left(\begin{matrix} 0 & ~~~~2j\\ 0 & ~~2p+1 \end{matrix} \right),\\\nonumber
B^{(3)}&=&\left(\begin{matrix}0 & ~~~~2j+1\\ 0 & ~~2p+1 \end{matrix} \right),\quad \quad B^{(4)}=\left(\begin{matrix} 0 & ~~~~2j+1\\ 0 & ~~2p \end{matrix} \right),\qquad 0 \leq j < k,\quad j,k,p\in \mathbb{Z}.
\end{eqnarray}
\normalsize
The sectors $B^{(2)}$, $B^{(3)}$ and $B^{(4)}$ in \eqref{8degsect} generate the orbifold shifts which mix the $SO(8)_1$ and $SO(8)_2$ with $SO(8)_3$ and $SO(8)_4$ which arise from the decomposition of a different $E_8$. The elliptic genus for the degenerate orbit is 
\small
\begin{eqnarray}
\label{8egdeg}
&&A_{\text{degenerate}}(q,R,F) =\\\nonumber
&&\frac{trR^4}{(2\pi)^4} \frac{E_4}{2^7 3^2 5 \eta^{24}}\left( B^{(1)} \sum_{a=2}^4 \theta^{16}_a + \sum_{j=2}^4 B^{(j)}2\theta_3^8 \theta_4^8\right) + \frac{(trR^2)^2}{(2\pi)^4} \frac{\et^2}{2^9 3^2  \eta^{24}}\left( B^{(1)} \sum_{a=2}^4 \theta^{16}_a +\sum_{j=2}^4 B^{(j)} 2\theta_3^8 \theta_4^8\right)\\\nonumber
&&- \frac{trR^2 \sum_{i=1}^4 trF_i^2}{2^8 3^2(2\pi)^4  \eta^{24}} \lbrace 2B^{(1)} (\et E_4 E_6-\et^2 E_4^2) -\sum_{j=2}^4 B^{(j)}(\et\theta_3^8 \theta_4^8 )(f_2 + f_3+2\et)\rbrace
\\\nonumber
&&+\frac{\sum_{i=1}^4 trF_i^4}{2^7 3 (2\pi)^4 \eta^{24}}\lbrace B^{(1)} (-\theta_2^{16}\theta_3^4 \theta_4^4 +\theta_3^{16}\theta_2^4 \theta_4^4 -\theta_4^{16}\theta_2^4 \theta_3^4) + \sum_{j=2}^4 B^{(j)}(\theta_3^8 \theta_4^8(\theta_2^4 \theta_4^4-\theta_2^4 \theta_3^4))\rbrace\\\nonumber
&&+ \frac{\sum_{i=1}^4(trF_i^2)^2 }{2^9 3^2 (2\pi)^4 \eta^{24}}\lbrace B^{(1)}\sum_{a=2}^4 \theta^{16}_a (\et+f_{a-1})^2 +\sum_{j=2}^4 B^{(j)}\theta_3^8 \theta_4^8 \left[(f_2 +\et)^2 +(f_3+\et)^2 \right]  \rbrace \\\nonumber
&&+\frac{trF_1^2 trF_2^2 + trF_3^2 trF_4^2}{2^8 3^2 (2\pi)^4 \eta^{24}}\lbrace B^{(1)}\sum_{a=2}^4 \theta^{16}_a (\et+f_{a-1})^2 +B^{(2)}\theta_3^8 \theta_4^8 \left[(f_2 +\et)^2 +(f_3+\et)^2 \right]\\\nonumber
&&\qquad \qquad \qquad \qquad \qquad \qquad +\sum_{j=3}^4 B^{(j)}\left[(f_2 +\et)^2 +(f_3+\et)^2 -9\theta_2^8\right] \rbrace\\\nonumber
&&+\frac{trF_1^2 trF_3^2 + trF_2^2 trF_4^2}{2^8 3^2 (2\pi)^4 \eta^{24}}\lbrace B^{(1)}\sum_{a=2}^4 \theta^{16}_a (\et+f_{a-1})^2 +B^{(3)}\theta_3^8 \theta_4^8 \left[(f_2 +\et)^2 +(f_3+\et)^2 \right]\\\nonumber
& &\qquad \qquad \qquad \qquad \qquad \qquad+( B^{(2)}+B^{(4)}) \left[(f_2 +\et)^2 +(f_3+\et)^2 -9\theta_2^8\right] \rbrace\\\nonumber
&&+\frac{trF_1^2 trF_4^2 + trF_2^2 trF_3^2}{2^8 3^2 (2\pi)^4 \eta^{24}}\lbrace B^{(1)}\sum_{a=2}^4 \theta^{16}_a (\et+f_{a-1})^2 +B^{(4)}\theta_3^8 \theta_4^8 \left[(f_2 +\et)^2 +(f_3+\et)^2 \right]\\\nonumber
&&\qquad \qquad \qquad \qquad \qquad \qquad +( B^{(2)}+B^{(3)}) \left[(f_2 +\et)^2 +(f_3+\et)^2 -9\theta_2^8\right] \rbrace.
\end{eqnarray}
\normalsize
Once again in the above, the $B^{(i)}$s remind one to take into account the different sectors as in \eqref{8degsect} while performing the final integration in \eqref{8int}.\\
Finally for the non-degenerate orbit ($B\neq 0$, $det(B)\neq 0$), we have to use the following sectors \cite{Gutperle:1999xu}
\small
\begin{align}
\label{8ndegsect}
B^{(1)}&=\left(\begin{matrix} 2k & ~2j\\ 0 & 2p \end{matrix} \right),\\\nonumber
 B^{(2,1)}=\left(\begin{matrix} 2k & ~2j\\ 0 & 2p+1 \end{matrix} \right),\quad  &B^{(2,2)}=\left(\begin{matrix} 2k & ~2j+1\\ 0 & 2p+1 \end{matrix} \right),\quad  B^{(2,3)}=\left(\begin{matrix} 2k & ~2j+1\\ 0 & 2p \end{matrix} \right)\\\nonumber
B^{(3)}=\left(\begin{matrix} 2k+1 & ~2j\\ 0 & 2p \end{matrix} \right),\quad \quad & B^{(4)}=\left(\begin{matrix} 2k+1 & ~2j+1\\ 0 & 2p \end{matrix} \right),\qquad 0 \leq j < k,\quad j,k,p\in \mathbb{Z}.
\end{align}
\normalsize
The sector $B^{(2)}$ has been divided in 3 subsectors $ B^{(2,1)},  ~B^{(2,2)},  ~B^{(2,3)}$ because of the spin structure in the elliptic genus \eqref{8eg}. To shorten the notation we shall use 
\begin{equation}
 B^{(2)}= B^{(2,1)}+ B^{(2,2)}+ B^{(2,3)},
\end{equation}
in the elliptic genus for the non-degenerate orbit (below) whenever the modular coefficients in front of $ B^{(2,k)}, k=1,2,3$ are same. We finally get the following elliptic genus for the non-degenerate orbit
\small
\begin{eqnarray}
\label{8egndeg}
&& A_{\text{non-degenerate}}(q,R,F) = \frac{trR^4}{(2\pi)^4} \frac{E_4}{2^7 3^2 5 \eta^{24}}\lbrace  B^{(1)} \sum_{a=2}^4 \theta^{16}_a +B^{(2)}2\theta_3^8 \theta_4^8 + B^{(3)}2\theta_2^8 \theta_3^8 +B^{(4)}2\theta_2^8 \theta_4^8 \rbrace \\\nonumber
&&  +\frac{(trR^2)^2}{(2\pi)^4} \frac{\et^2}{2^9 3^2  \eta^{24}}\lbrace B^{(1)} \sum_{a=2}^4 \theta^{16}_a +B^{(2)}2\theta_3^8 \theta_4^8 + B^{(3)}2\theta_2^8 \theta_3^8 +B^{(4)}2\theta_2^8 \theta_4^8 \rbrace\\\nonumber
&&- \frac{trR^2 \sum_{i=1}^4 trF_i^2}{2^8 3^2(2\pi)^4  \eta^{24}} \lbrace(  2B^{(1)} (\et E_4 E_6-\et^2 E_4^2) -B^{(2)}(\et\theta_3^8 \theta_4^8 )(f_2 + f_3+2\et)\\\nonumber
&&\qquad \qquad \qquad \qquad -B^{(3)}(\et\theta_2^8 \theta_3^8 )(f_1 + f_2+2\et)-B^{(4)}(\et\theta_2^8 \theta_4^8 )(f_1 + f_3+2\et)\rbrace \\\nonumber
&&+\frac{\sum_{i=1}^4 trF_i^4}{2^7 3 (2\pi)^4 \eta^{24}} \lbrace B^{(1)} (-\theta_2^{16}\theta_3^4 \theta_4^4 +\theta_3^{16}\theta_2^4 \theta_4^4 -\theta_4^{16}\theta_2^4 \theta_3^4) + B^{(2)}(\theta_3^8 \theta_4^8(\theta_2^4 \theta_4^4-\theta_2^4 \theta_3^4))\\\nonumber
&&\qquad \qquad \qquad \qquad + B^{(3)}(\theta_2^8 \theta_3^8(\theta_2^4 \theta_4^4-\theta_3^4 \theta_4^4))+ B^{(4)}(\theta_2^8 \theta_4^8(-\theta_3^4 \theta_4^4-\theta_2^4 \theta_3^4))\rbrace\\\nonumber
&&+ \frac{\sum_{i=1}^4(trF_i^2)^2}{2^9 3^2 (2\pi)^4 \eta^{24}}\lbrace B^{(1)}\sum_{a=2}^4 \theta^{16}_a (\et+f_{a-1})^2 +B^{(2)}\theta_3^8 \theta_4^8 \left[(f_2 +\et)^2 +(f_3+\et)^2 \right]\\\nonumber
&&\qquad \qquad \qquad \qquad  +B^{(3)}\theta_2^8 \theta_3^8 \left[(f_1 +\et)^2 +(f_2+\et)^2 \right] +B^{(4)}\theta_2^8 \theta_4^8 \left[(f_1 +\et)^2 +(f_3+\et)^2 \right]\rbrace \\\nonumber
&&+\frac{trF_1^2 trF_2^2 + trF_3^2 trF_4^2}{2^8 3^2 (2\pi)^4 \eta^{24}}\lbrace B^{(1)}\sum_{a=2}^4 \theta^{16}_a (\et+f_{a-1})^2 +B^{(2)}\theta_3^8 \theta_4^8 \left[(f_2 +\et)^2 +(f_3+\et)^2 \right]\\\nonumber
&&\qquad \qquad \qquad \qquad \qquad \qquad -(B^{(2,2)} +B^{(2,3)})2^8 3^2 \eta^{24} + B^{(3)}\theta_2^8 \theta_3^8 \left[(f_1 +\et)^2 +(f_3+\et)^2 \right]\\\nn
&& \qquad \qquad \qquad \qquad \qquad \qquad +B^{(4)}\theta_2^8 \theta_4^8 \left[(f_1 +\et)^2 +(f_3+\et)^2 \right]\rbrace\\\nonumber
&&+\frac{trF_1^2 trF_3^2 + trF_2^2 trF_4^2}{2^8 3^2 (2\pi)^4 \eta^{24}}\lbrace B^{(1)}\sum_{a=2}^4 \theta^{16}_a (\et+f_{a-1})^2 +B^{(2)}\theta_3^8 \theta_4^8 \left[(f_2 +\et)^2 +(f_3+\et)^2 \right]\\\nn
&&  \qquad \qquad \qquad \qquad \qquad \qquad  -(B^{(2,1)}+B^{(2,3)})2^8 3^2 \eta^{24}+ B^{(3)}\theta_2^8 \theta_3^8 \left[2(f_1 +\et)(f_3+\et) \right]\\\nonumber
&&\qquad \qquad \qquad \qquad \qquad \qquad+B^{(4)}\theta_2^8 \theta_4^8 \left[2(f_1 +\et)(f_3+\et) \right]\rbrace\\\nonumber
&&+\frac{trF_1^2 trF_4^2 + trF_2^2 trF_3^2}{2^8 3^2 (2\pi)^4 \eta^{24}}\lbrace B^{(1)}\sum_{a=2}^4 \theta^{16}_a (\et+f_{a-1})^2 +B^{(2)}\theta_3^8 \theta_4^8 \left[(f_2 +\et)^2 +(f_3+\et)^2 \right]\\\nn
&&  \qquad \qquad \qquad \qquad \qquad \qquad  -(B^{(2,1)}+B^{(2,2)})2^8 3^2 \eta^{24}+ B^{(3)}\theta_2^8 \theta_3^8 \left[2(f_1 +\et)(f_3+\et) \right]\\\nonumber
&&\qquad \qquad \qquad \qquad  \qquad \qquad +B^{(4)}\theta_2^8 \theta_4^8 \left[2(f_1 +\et)(f_3+\et) \right]\rbrace.
\end{eqnarray}
\normalsize
The complete CP-even amplitude will be (see \eqref{int} for the integration domains)
\small
\begin{eqnarray}
  \label{int3}
  \mathcal{A} &=& \mathcal{A}_{\text{trivial}}+\mathcal{A}_{\text{degenerate}}+\mathcal{A}_{\text{non-degenerate}}
  \end{eqnarray}
  \normalsize
  with 
  \small
   \begin{eqnarray}
  \label{8trivamp}
&&  \mathcal{A}_{\text{trivial}}= T_2 V_8 t_8\lbrace trR^4+\frac{1}{4}(trR^2)^2+trR^2 \sum_{i=1}^4 trF_i^2\\\nn
&&-2trF_1^2trF_3^2-2trF_1^2trF_4^2-2trF_2^2trF_4^2-2trF_1^2trF_3^2 +4trF_1^2trF_2^2+4trF_3^2trF_4^2 +2\sum_{i=1}^4(trF_i^2)^2\rbrace
  \end{eqnarray}
  \normalsize
  being the trivial orbit amplitude. Note that by recombining the $SO(8)_1$ with $SO(8)_2$ and $SO(8)_3$ with $SO(8)_4$ we find back the $T^2$ reduction of the $\ett$ Green-Schwarz term \eqref{e88gs}.\\
  We now collect the constant parts of the q-expansion of the modular functions in the degenerate and non-degenerate elliptic genus from \eqref{const}, verify that there are no poles and then use the integral \eqref{degint} to evaluate the non-volume suppressed harmonic part of the degenerate amplitude and \eqref{ndegint} to evaluate the non-volume suppressed harmonic part of the the non-degenerate amplitude in the CP-even sector. 
  \small
  \begin{eqnarray}
  \label{8degamp}
  && \mathcal{A}_{\text{degenerate}}=\\\nn
  &&\frac{1}{4 (2\pi)^4} \leemu V_8 t_8\left[ trR^4+\frac{1}{4}(trR^2)^2+trR^2 \sum_i^4 trF_i^2+\sum_i^4 2(trF_i^2)^2\right]\\\nn
  && +\frac{1}{4 (2\pi)^4}\leemu \times\\\nn
  && \quad  V_8 t_8\left[ 2 \times 64 \left( \frac{trR^4}{360} + \frac{(trR^2)^2}{288}\right)+\frac{4}{3}trR^2 \sum_{i=1}^4 trF_i^2 + 2 \times\frac{2}{3} \left(4\sum_{i=1}^4 trF_i^4 +3trF_1^2 trF_2^2 +3 trF_3^2 trF_4^2  \right)  \right]\\\nn
  && + \frac{1}{8 (2\pi)^4}\leemu \times V_8 t_8\lbrace 256 \left( \frac{trR^4}{360} + \frac{(trR^2)^2}{288}\right)+\frac{8}{3}trR^2 \sum_{i=1}^4 trF_i^2\\\nn
   &&\qquad \qquad \qquad \qquad \qquad \qquad  + \frac{8}{3} \left(-4\sum_{i=1}^4 trF_i^4 +3\sum_{i=1}^4 (trF_i^2)^2 +6trF_1^2 trF_2^2 +6 trF_3^2 trF_4^2  \right)\rbrace\\\nn
   &&+\frac{1}{(2\pi)^4} \leemu V_8 t_8\left[ trF_1^2 trF_2^2+trF_3^2 trF_4^2 +trF_1^2 trF_3^2+trF_2^2 trF_4^2+trF_1^2 trF_4^2+trF_2^2 trF_3^2\right].
  \end{eqnarray}
  \normalsize
  \small
  \begin{eqnarray}
  \label{8ndegamp}
  && \mathcal{A}_{\text{non-degenerate}}=\\\nn
  &&\frac{1}{4(2\pi)^4}\left[ \frac{\pi T_2}{3}+ln \left( T_2|\eta(2T)|^2\right) \right]  V_8 t_8\left[ trR^4+\frac{1}{4}(trR^2)^2+trR^2 \sum_i^4 trF_i^2+\sum_i^4 2(trF_i^2)^2\right]\\\nn
  && +\frac{1}{4(2\pi)^4}\left[ \frac{\pi T_2}{3}+ln \left( T_2|\eta(2T)|^2\right) \right]   \times\\\nn
  && \quad V_8 t_8 \left[ 2 \times 64 \left( \frac{trR^4}{360} + \frac{(trR^2)^2}{288}\right)+\frac{4}{3}trR^2 \sum_{i=1}^4 trF_i^2 + 2 \times\frac{2}{3} \left(4\sum_{i=1}^4 trF_i^4 +3trF_1^2 trF_2^2 +3 trF_3^2 trF_4^2  \right)  \right]\\\nn
  && + \frac{1}{8 (2\pi)^4}\left[\frac{\pi T_2}{3} +ln \left( T_2|\eta(4T)|^2\right) -ln \left( T_2|\eta(2T)|^2\right)  \right]  \times V_8 t_8\lbrace 256 \left( \frac{trR^4}{360} + \frac{(trR^2)^2}{288}\right)+\frac{8}{3}trR^2 \sum_{i=1}^4 trF_i^2  \\\nn
   && \qquad \qquad \qquad \qquad \qquad \qquad \qquad \qquad \qquad +\frac{8}{3} \left(-4\sum_{i=1}^4 trF_i^4 +3\sum_{i=1}^4 (trF_i^2)^2 +6trF_1^2 trF_2^2 +6 trF_3^2 trF_4^2  \right)\rbrace\\\nn
   &&+\frac{1}{ (2\pi)^4} \left[ \frac{\pi T_2}{3}+ln \left( T_2|\eta(2T)|^2\right) \right]   V_8 t_8 \left[ trF_1^2 trF_2^2+trF_3^2 trF_4^2 \right]\\\nn
   &&+\frac{1}{ (2\pi)^4}\left[\frac{\pi T_2}{3} +ln \left( T_2|\eta(4T)|^2\right) -ln \left( T_2|\eta(2T)|^2\right)  \right]\times V_8 t_8\left[ trF_1^2 trF_3^2+trF_2^2 trF_4^2\right] \\\nn
   &&+\frac{1}{ (2\pi)^4}\left[ ln \left(T_2 |\eta(4T)|^2\right)-2ln \left(T_2 |\eta(2T)|^2\right)  \right]\times V_8 t_8\left[ trF_1^2 trF_4^2+trF_2^2 trF_3^2\right]  .
  \end{eqnarray}
  \normalsize
  
  We note that the 8-form polynomial 
  \begin{equation}
  \label{so8ano}
  Y_8=\left[ trR^4+\frac{1}{4}(trR^2)^2+trR^2 \sum_i^4 trF_i^2+\sum_i^4 2(trF_i^2)^2\right]
  \end{equation} is due to the fermions transforming under the adjoint representation ${(28,1,1,1)}\oplus {(1,28,1,1)}\oplus {(1,1,28,1)}\oplus {(1,1,1,28)}$ of the $SO(8)^4$ and the CP-odd partner of the above provides with the discrete $\sz$ anomaly cancelling counter-term in D=8, N=1 supergravity with gauge group $G=SO(8)^4$ \cite{Minasian:2016hoh}. The other two 8-form polynomials
  \begin{eqnarray}
 && Y_8'=\lbrace 2 \times 64 \left( \frac{trR^4}{360} + \frac{(trR^2)^2}{288}\right)+\frac{4}{3}trR^2 \sum_{i=1}^4 trF_i^2 \\
 && \qquad \qquad \qquad \qquad \qquad + 2 \times\frac{2}{3} \left(4\sum_{i=1}^4 trF_i^4 +3trF_1^2 trF_2^2 +3 trF_3^2 trF_4^2  \right)  \rbrace\\
&&  Y_8''=\lbrace 256 \left( \frac{trR^4}{360} + \frac{(trR^2)^2}{288}\right)+\frac{8}{3}trR^2 \sum_{i=1}^4 trF_i^2  \\\nn
   && \qquad \qquad \qquad \qquad \qquad +\frac{8}{3} \left(-4\sum_{i=1}^4 trF_i^4 +3\sum_{i=1}^4 (trF_i^2)^2 +6trF_1^2 trF_2^2 +6 trF_3^2 trF_4^2  \right)\rbrace
  \end{eqnarray}
are respectively the contributions from the massive vector multiplet transforming under the bi-fundamental representations $(8,8,1,1)\oplus (1,1,8,8)$ and (co)spinor representations $(8,8,1,1)'\oplus (1,1,8,8)'$ respectively. The last few pure gauge terms in \eqref{8ndegamp} are due to the orbifold shifts \cite{Lerche:1998nx, Kiritsis:2000zi}.
%============================================================================
\subsection{Calculating string amplitude with Hecke operators}\label{sec:hecke}
We now deploy the elegant method of Hecke operator to evaluate the degenerate  plus non-degenerate CP-even amplitude $ \mathcal{A}_{\text{degenerate}}+\mathcal{A}_{\text{non-degenerate}}$ which have been carried out in \cite{Kiritsis:2000zi} and in the guise of modular identities in \cite{Lerche:1998nx}. We complement the calculation of \cite{Kiritsis:2000zi} where only the $\Gamma_2^-$ subgroup (of $SL(2,\mathbb{Z})$) invariant part has been computed using the Hecke image $\Heckea$ of the $\Ga$ invariant part of $ \mathcal{A}_{\text{degenerate}}+\mathcal{A}_{\text{non-degenerate}}$ in \eqref{8degamp} and \eqref{8ndegamp}. We compute the $\Gb$ and $\Gc$ invariant parts of \eqref{8degamp} and \eqref{8ndegamp}
 using the method of Hecke operators $\Heckeb$ and $\Heckec$. We shall see that in the pure gravitational and in mixed gauge gravity part we can separate the contribution from the adjoint representation \eqref{so8ano} and the total contribution from 6 sets of bi-fundamental states like $(8,8,1,1)$ etc. but in the pure gauge part we cannot separate these contributions: instead the sum from the 3 subgroups $\Ga$, $\Gb$ and $\Gc$ of $SL(2,\mathbb{Z})$ we shall retrieve the total pure gauge contributions which have been investigated in detail in \cite{Lerche:1998nx}.\\
We now describe the method in brief. For exclusive details we refer to \cite{Kiritsis:2000zi}. We note that subgroups $\Ga$, $\Gb$ and $\Gc$ are the invariant subgroups of $\theta_2$, $\theta_4$ and $\theta_3$ modulo the phase and weight factors. Now using the \eqref{fsum} and \eqref{sumid1}, \eqref{sumid2}, \eqref{sumid2}, \eqref{sumid4} summation identities we can decompose the $B^{(1)}$ part in both degenerate and non-degenerate elliptic genus \eqref{8egdeg} and \eqref{8egndeg} into sum of the form
\begin{equation}
\label{a1}
B^{(1)}(\cdots)= B^{(1)}\theta_3^8 \theta_4^8 (\cdots) +  B^{(1)}\theta_2^8 \theta_3^8 (\cdots)+ B^{(1)}\theta_2^8 \theta_4^8 (\cdots).
\end{equation}
One can now combine the part $B^{(1)}\theta_3^8 \theta_4^8 (\cdots)$ with $B^{(2)}$, $B^{(3)}$ and $B^{(4)}$ sectors \eqref{8degsect} in the degenerate elliptic genus \eqref{8egdeg} and  $B^{(2,1)}$, $B^{(2,2)}$ and $B^{(2,3)}$ sectors \eqref{8ndegsect} in the non-degenerate elliptic genus \eqref{8egndeg}. The sum over $\theta_3^8 \theta_4^8 (\cdots)$ is then of the form
\begin{equation}
\label{-int1}
\int_{\mathcal{F}^-}\frac{d^2 \tau}{\tau_2^2} \Gamma_{2,2} (2T, U;2\tau) \Phi_{-}(\tau)
\end{equation}
where $ \Phi_{-}(\tau)$ is $\Ga$ invariant modular function and we restrict the integral domain to $\mathcal{F}^-$ which is the fundamental domain of $\Ga$ subgroup. One can now change the variable $2\tau = \rho$ and unfold the integral \eqref{-int1} to the fundamental domain $\mathcal{F}$ of $SL(2,\mathbb{Z})$ by the following unfolding
\small
\begin{eqnarray}
\label{-int2}
& &\int_{\mathcal{F}^-}\frac{d^2 \tau}{\tau_2^2} \Gamma_{2,2} (2T, U;2\tau) \Phi_{-}(\tau)=\int_{\mathcal{F}}\frac{d^2 \rho}{\rho_2^2} \Gamma_{2,2} (2T, U;\rho) \left( \Phi_{-}(\frac{\rho}{2})+\Phi_{-}(-\frac{1}{2\rho}) + \Phi_{-}(\frac{\rho+1}{2}) \right) \\\nonumber
& &=\int_{\mathcal{F}}\frac{d^2 \rho}{\rho_2^2} \Gamma_{2,2} (2T, U;\rho) \Heckea \Phi_{-}(\rho)
\end{eqnarray}
\normalsize
where in the last line we have used the definition of the Hecke operator for the $\Ga$ subgroup.\\
We then combine the $\theta_2^8 \theta_3^8 (\cdots)$ piece in \eqref{a1} with the $B^{(3)}$ sector of the non-degenerate elliptic genus to get the following combination of the partitions function
\begin{equation}
\label{+int1}
\int_{\mathcal{F}^+}\frac{d^2 \tau}{\tau_2^2} \Gamma_{2,2} (2T, U;\tau/2) \Phi_{+}(\tau)
\end{equation}
where $ \Phi_{+}(\tau)$ is $\Gb$ invariant modular function and we restrict the integral domain to $\mathcal{F}^+$ which is the fundamental domain of $\Gb$ subgroup. We make the change of variable $\tau/2 = \rho$ and unfold the integral to the fundamental domain $\mathcal{F}$ to make appear the Hecke operator $\Heckeb$ for the $\Gb$ subgroup
\small
\begin{eqnarray}
\label{+int2}
& &\int_{\mathcal{F}^+}\frac{d^2 \tau}{\tau_2^2} \Gamma_{2,2} (2T, U;\tau/2) \Phi_{+}(\tau)=\int_{\mathcal{F}}\frac{d^2 \rho}{\rho_2^2} \Gamma_{2,2} (2T, U;\rho) \left( \Phi_{+}(2\rho)+\Phi_{+}(-\frac{1}{2\rho}) + \Phi_{+}(\frac{1}{2\rho+1}) \right) \\\nonumber
& &=\int_{\mathcal{F}}\frac{d^2 \rho}{\rho_2^2} \Gamma_{2,2} (2T, U;\rho) \Heckeb \Phi_{+}(\rho).
\end{eqnarray}
\normalsize
 It now rests to combine the $\theta_2^8 \theta_4^8 (\cdots)$ piece in \eqref{a1} with the $B^{(4)}$ sector of the non-degenerate elliptic genus to get the following combination of the partitions function
\begin{equation}
\label{0int1}
\int_{\mathcal{F}^0}\frac{d^2 \tau}{\tau_2^2} \Gamma_{2,2} (T, U;(\tau+1)/2) \Phi_{0}(\tau)
\end{equation}
where $ \Phi_{0}(\tau)$ is $\Gc$ invariant modular function and we restrict the integral domain to $\mathcal{F}^0$ which is the fundamental domain of $\Gc$ subgroup. Making the change of variable $(\tau+1)/2 = \rho$ and unfold the integral to the fundamental domain $\mathcal{F}$ to make appear the Hecke operator $\Heckec$ for the $\Gc$ subgroup
\small
\begin{eqnarray}
\label{0int2}
& &\int_{\mathcal{F}^0}\frac{d^2 \tau}{\tau_2^2} \Gamma_{2,2} (T, U;(\tau+1)/2) \Phi_{0}(\tau)=\int_{\mathcal{F}}\frac{d^2 \rho}{\rho_2^2} \Gamma_{2,2} (T, U;\rho) \Heckec \Phi_{0}(\rho)\\\nonumber
& &=\int_{\mathcal{F}}\frac{d^2 \rho}{\rho_2^2} \Gamma_{2,2} (T, U;\rho) \left( \Phi_{0}(2\rho-1)+\Phi_{0}(-\frac{1}{2\rho-1}) + \Phi_{0}(-\frac{1}{2\rho}) \right) .
\end{eqnarray}
\normalsize
Now to get the harmonic part of the CP-even amplitude, we pick up the constant parts of the Hecke images of the related modular functions which we have enlisted in \eqref{f4sum}, \eqref{hecke1}, \eqref{hecke2} and \eqref{hecke3}. Combining these we find the result for degenerate and non-degenerate amplitude
\small
\begin{eqnarray}
\label{h}
& &\mathcal{A}_{\text{degenerate}}+\mathcal{A}_{\text{non-degenerate}}=\\\nonumber
&  & \frac{1}{(2\pi)^4} \left[ \frac{\pi T_2}{3} +ln T_2|\eta(2T)|^2 +\leemu\right]t_8\underbrace{\left[ \frac{2 \times 360}{2^7 3^2 5} trR^4+\frac{2 \times 72}{2^9 3^2 }(trR^2)^2+\frac{288}{2^8 3^2} trR^2 \sum_i^4 trF_i^2\right]}_{\text{adjoint of }SO(8)^4}\\\nonumber
&  &+\frac{1}{(2\pi)^4}\left[ \frac{\pi T_2}{2}+ln T_2|\eta(2T)|^2+\leemt +\leemu\right]\\\nonumber
& &\times t_8\underbrace{ \left[ \frac{2 \times 384}{2^7 3^2 5} trR^4+\frac{2 \times 384}{2^9 3^2 }(trR^2)^2+\frac{1152}{2^8 3^2} trR^2 \sum_i^4 trF_i^2\right]}_{\text{bi-fundamental and bi-spinor states}}\\\nonumber
& & + \frac{1}{(2\pi)^4} \left[ln T_2|\eta(2T)|^2-2\leemt +\leemu\right]t_8 \sum_{i=1}^4(trF_i^2)^2\\\nonumber
&  &+\frac{1}{ (2\pi)^4} \left[\frac{\pi T_2}{3} +lnT_2 |\eta(4T)|^2-lnT_2|\eta(2T)|^2 \right] t_8\sum_{i=1}^4 trF_i^4\\
& & +\frac{1}{(2\pi)^4}\left[ \frac{\pi T_2}{3} +ln T_2|\eta(4T)|^2-lnT_2|\eta(2T)|^2 +\leemu\right]\times t_8\{trF_1^2 trF_3^2+trF_2^2 trF_4^2\}\\
& & +\frac{1}{(2\pi)^4}\left[ ln T_2|\eta(4T)|^2-2lnT_2|\eta(2T)|^2  +\leemu\right]\times t_8\{trF_1^2 trF_4^2+trF_2^2 trF_3^2\}.
\end{eqnarray}
\normalsize
From the above, we recognise the composite anomaly cancelling polynomial \eqref{so8ano} \cite{Minasian:2016hoh} in pure gravity and gauge-gravity sector in part ``adjoint of $SO(8)^4$" \eqref{h} and the part ``bi-fundamental and bi-spinor states in" corresponds to the pure gravity and gauge-gravity coupling of states in $(8,8,1,1)\oplus (1,1,8,8)$, $(8,8,1,1)'\oplus (1,1,8,8)'$ and $(8,8,1,1)''\oplus (1,1,8,8)''$ representations. However the pure gauge sector irons down the contributions from these representations to give the last terms in \eqref{h}. One can also check that there is a ``local conservation" of coefficients e.g. for $trR^4/(2^7 3^2 5)$ terms in both methods with constant coefficients and Hecke operators the total numerical coefficients are same if one sums them in the respective sectors 
\begin{equation}
2\times 744 = \underbrace{2 \times 360 + 2 \times 384 }_{\text{Hecke metheod}} = \underbrace{2 \times 360 +2 \times 128 + 2 \times 256}_{\text{adjpoint + bi-fundamental + spinor reps.}}.
\end{equation}
One can check the other numerical coefficients for the 8-forms $(trR^2)^2$, $(trF^2)^2$, $trR^2 trF^2$ and $trF^4$. There is a nice interpretation for the modular forms in front of the pure gauge sector 8-forms as discussed in \cite{Lerche:1998nx} and they correspond to the $C_4$ and $C_0 -C_8$ exchange between four $\mathcal{D}_4$ branes in the dual F-theory on K3 description in Sen limit \cite{Sen:1996vd}.

%===============================================================
\section{Conclusion}
We have summarized the one loop amplitude results for Heterotic string on $T^2$ with gauge groups $SO(32)$, $\ett$, $\sx$ and $SO(8)^4$. We have emphasized the role played by the harmonic part of these amplitudes which provide the discrete anomaly counter-term in the corresponding supergravity theory. We also discussed instances where these terms have an uplift towards nine-dimensions thus providing a consistent description of both the amplitude calculation for a circle compactification and for a torus compactification. For the case of $SO(8)^4$ we complemented the calculation of such amplitude using the Hecke operators which may be seen to provide an interesting perspective towards such calculations from number theory point of view. We shall, in future, address the calculation of two and higher loop amplitudes for Heterotic string on $T^2$, the results of which provide higher derivative correction terms to the corresponding supergravity actions and are still not extensively studied in the string theory literature.

%%%%%%%%%%%%%%%%%%%%%%%%%%%%%%%%%%%%%%%%%%%%%%%%
\section*{Acknowledgements}
%%%%%%%%%%%%%%%%%%%%%%%%%%%%%%%%%%%%%%%%%%%%%%%%

We would like to thank  Pierre Vanhove and Boris Pioline for helpful discussions and valuable insights.  We would also like to thank Ruben Minasian and Raffaele Savelli for their support and collaboration in course of this work.

%========================================================================
\newpage
\appendix

%========================================================================
\section{Modular functions}\label{sec: modfunc}

In this appendix we provide the definitions the the Jacobi $\theta$ functions, Dedekind eta function and Eisenstein series along with useful identities relating them that we have used in the calculations.\\
Our convention for the $\theta$ function is
\begin{equation}
\theta\left[ \substack{a \\ b}\right](\nu|\tau)=\sum_{n\in \mathbb{Z}}q^{(1/2)(n-a/2)^2}e^{2\pi i (\nu-b/2)(n-a/2)},
\end{equation}
where a,b are real and $q=e^{2\pi i \tau}$.\\
We note $\theta_1=\theta\left[ \substack{1 \\ 1}\right]$, $\theta_2=\theta\left[ \substack{1 \\ 0}\right]$, $\theta_3=\theta\left[ \substack{0 \\ 0}\right]$ and $\theta_4=\theta\left[ \substack{0 \\ 1}\right]$.\\
Next we list different periodicity properties and modular transformations of the $\theta$ functions ($a,b \in\mathbb{Z}$):
\begin{eqnarray}
\label{thetaperiod}
\theta\left[ \substack{a+2 \\ b}\right](\nu|\tau)&=&\theta\left[ \substack{a \\ b}\right](\nu|\tau),\\\nonumber
\theta\left[ \substack{a \\ b+2}\right](\nu|\tau)&=&e^{i\pi a}\theta\left[ \substack{a \\ b}\right](\nu|\tau),\\\nonumber
\theta\left[ \substack{-a \\ -b}\right](\nu|\tau)&=&\theta\left[ \substack{a \\ b}\right](-\nu|\tau),\\\nonumber
\theta\left[ \substack{a \\ b}\right](-\nu|\tau)&=&e^{i\pi ab}\theta\left[ \substack{a \\ b}\right](\nu|\tau),\\\nonumber
\theta\left[ \substack{a \\ b}\right](\nu|\tau+1)&=&e^{-(i\pi/4)a(a-2)}\theta\left[ \substack{a \\ a+b-1}\right](\nu|\tau),\\\nonumber
\theta\left[ \substack{a \\ b}\right](\nu/\tau|-1/\tau)&=&\sqrt{-i\tau}e^{(i\pi/2)ab+i\pi\nu^2/\tau}\theta\left[ \substack{b \\ -a}\right](\nu|\tau).\nonumber
\end{eqnarray}
We are now in position to define the Dedekind $\eta$-function:
\begin{equation}
\eta(\tau)=q^{1/24}\prod_{n=1}^{\infty}(1-q^n),
\end{equation}
satisfying \begin{equation}
\eta(-1/\tau)=\sqrt{-i\tau}\eta(\tau).
\end{equation}
Some useful relations between the Jacobi $\theta$-functions and the $\eta$-function are
\begin{subequations}
\label{th-eta}
\begin{align}
\theta_2(0|\tau) \theta_3(0|\tau) \theta_4(0|\tau) &=2 \eta^3,\\
\theta^{12}_3 - \theta^{12}_2 -\theta^{12}_4&=48 \eta^{12},\\
\theta^{4}_2+\theta^{4}_4-\theta^{4}_3=0.
\end{align}
\end{subequations}
%=====================================================================
Now we summarise the definitions of the Eisenstein series and Leech $j$ function
\begin{subequations}
\label{eisenstein}
\begin{align}
  \hat{E}_2&=1-\frac{3}{\pi \tau_2}-24 \sum_{n=1}^{\infty}\frac{nq^n}{1-q^n},\\
  E_4&=\frac{1}{2}\sum_{a=2}^4 \theta_a^8=1+240\sum_{n=1}^{\infty}\frac{n^3q^n}{1-q^n},\\
  E_8&=E_4^2=\frac{1}{2}\sum_{a=2}^4 \theta_a^{16}=1+480\sum_{n=1}^{\infty}\frac{n^7q^n}{1-q^n},\\
  E_6&=\frac{1}{2}(\theta^4_2+\theta^4_3)(\theta^4_3+\theta^4_4)(\theta^4_4-\theta^4_2)=1-504\sum_{n=1}^{\infty}\frac{n^5q^n}{1-q^n},\\
   {j}&= \frac{{E}^3_4}{{\eta}^{24}}=\frac{1}{{q}}+744+\cdots
    \end{align}
  \end{subequations}
  
  In the process of ``gauging" the elliptic genus, we shall extensively use the following identities
\begin{subequations}  
\label{thetapartition}
\begin{align}
\frac{\theta_2(\nu|\tau)}{\theta_2(0|\tau)}&=exp\left\lbrace  \sum_{k=1}^{\infty}\frac{(2\pi i)^{2k} B_{2k}\nu^{2k}}{(2k+1)!-(2k)!} \left[E_{2k}(q)-2^{2k}E_{2k}(q^2) \right]  \right\rbrace \\
\frac{\theta_3(\nu|\tau)}{\theta_3(0|\tau)}&=exp\left\lbrace  \sum_{k=1}^{\infty}\frac{(2\pi i)^{2k} B_{2k}\nu^{2k}}{(2k+1)!-(2k)!} \left[E_{2k}(q)-E_{2k}(-\sqrt{q}) \right]  \right\rbrace \\
\frac{\theta_4(\nu|\tau)}{\theta_4(0|\tau)}&=exp\left\lbrace  \sum_{k=1}^{\infty}\frac{(2\pi i)^{2k} B_{2k}\nu^{2k}}{(2k+1)!-(2k)!} \left[E_{2k}(q)-E_{2k}(\sqrt{q}) \right]  \right\rbrace 
\end{align}
  \end{subequations}
  where $B_k$ are the Bernoulli numbers: $B_2=1/6,~ B_4=-1/30,~B_6=1/42$ and we shall use the following combinations $f_1$, $f_2$, $f_3$ in the elliptic genus
\begin{subequations}  
\label{deff}
\begin{align}
f_1 &=4E_2(q^2)-2E_2(q)=\theta_3^4 + \theta_4^4,\\
f_2 &= E_2(-\sqrt{q})-2E_2(q)=\theta_2^4 - \theta_4^4,\\
f_3 &= E_2(\sqrt{q})-2E_2(q)=-\theta_2^4-\theta_3^4.
\end{align}
  \end{subequations}
  
  \begin{subequations} 
  \label{fpart} 
\begin{align}
E_4(q)-16E_4(q^2)&=5(E_4(q)^2-f_1^2)=-15\theta_3^4 \theta_4^4,\\
 E_4(q)-E_4(-\sqrt{q})&=5(E_4(q)^2-f_2^2)=15\theta_2^4 \theta_4^4,\\
 E_4(q)-E_4(\sqrt{q})&=5(E_4(q)^2-f_3^2)=-15\theta_2^4 \theta_3^4.
\end{align}
  \end{subequations}
  There are various summation identities involving the Eisenstein series and $f_1$, $f_2$, $f_3$ which will be useful in the computation of the partition function
 \begin{subequations}  
 \label{fsum}
\begin{align} 
\frac{1}{2}\sum_{a=2}^4 \theta_a(0|\tau)^{16}f_{a-1}&=-E_4 E_6,\\
\frac{1}{2}\sum_{a=2}^4 \theta_a(0|\tau)^{16}f_{a-1}^2&=E_4^3-2^7 3^2 \eta^{24}=2 E_6^2-E_4^3,\\
\frac{1}{2}\sum_{a=2}^4 \theta_a(0|\tau)^{8}f_{a-1}&=-E_6,\\
\frac{1}{2}\sum_{a=2}^4 \theta_a(0|\tau)^{8}f_{a-1}^2&=E_4^2,
\end{align}
  \end{subequations}
  
 \begin{subequations}  
 \label{sumid}
\begin{align} 
\sum_{a=2}^4 \theta_a(0|\tau)^{16}&=2\theta_3^8 \theta_4^8 +2\theta_2^8 \theta_4^8 +2\theta_2^8 \theta_3^8 ,\label{sumid1}\\
\sum_{a=2}^4 \theta_a(0|\tau)^{16} (\hat{E}_2 +f_{a-1}) &=\theta_3^8 \theta_4^8 (2\hat{E}_2 +f_2+f_3)\label{sumid2}\\\nonumber
{      }&+\theta_2^8 \theta_4^8 (2\hat{E}_2 +f_1+f_3)+\theta_2^8 \theta_3^8 (2\hat{E}_2 +f_1+f_2),\\
\sum_{a=2}^4 \theta_a(0|\tau)^{16} (\hat{E}_2 +f_{a-1})^2 &=2\theta_3^8 \theta_4^8 (\hat{E}_2 +f_2)(\hat{E}_2+f_3)\label{sumid3}\\\nonumber
{      }&{  }~~+2\theta_2^8 \theta_4^8 (\hat{E}_2 +f_1)(\hat{E}_2+f_3)\\\nonumber
{      }&{  }~~+2\theta_2^8 \theta_3^8 (\hat{E}_2 +f_1)(\hat{E}_2+f_2)+2^8 3^2 \eta^{24},\\
{         }&=\theta_3^8 \theta_4^8 \left( (\hat{E}_2 +f_2)^2+(\hat{E}_2+f_3)^2\right) \label{sumid4}\\\nonumber
{      }&{  }~~+\theta_2^8 \theta_4^8 \left( (\hat{E}_2 +f_1)^2+(\hat{E}_2+f_3)^2\right)\\\nonumber
{      }&{  }~~+\theta_2^8 \theta_3^8 \left( (\hat{E}_2 +f_1)^2+(\hat{E}_2+f_2)^2\right)-2^9 3^2 \eta^{24}.
\end{align}
  \end{subequations}
  For the pure gauge part there are very remarkable trivial identities
\begin{align}
\label{f4sum}
\frac{1}{2^8 3 \eta^{24}}\left( -\theta^{16}_2 \theta^4_3 \theta^4_4 + \theta^{16}_3 \theta^4_2 \theta^4_4  -\theta^{16}_4 \theta^4_2 \theta^4_3\right) &=1,\\\nonumber
\frac{\theta^8_2 \theta^8_3}{2^8 3 \eta^{24}}\left( - \theta^4_3 \theta^4_4 +  \theta^4_2 \theta^4_4  \right)&=-\frac{1}{3},\\\nonumber
\frac{\theta^8_3 \theta^8_4 }{2^8 3 \eta^{24}}\left( + \theta^4_2 \theta^4_4 - \theta^4_2 \theta^4_3  \right)&=-\frac{1}{3},\\\nonumber
\frac{\theta^8_2 \theta^8_4 }{2^8 3 \eta^{24}}\left( - \theta^4_3 \theta^4_4 -  \theta^4_2 \theta^4_3  \right)&=-\frac{1}{3}.\nonumber
\end{align}

Then we enlist the q-expansions of the different modular functions used in the elliptic genus
%\begin{eqnarray}  
\small
\begin{align}
\label{const}
\frac{E_4}{\eta^{24}}\sum_{a=2}^4 \theta_a(0|\tau)^{16} =\frac{2}{q}+1488+\mathcal{O}(q),&\qquad \qquad \frac{E_4}{\eta^{24}}\theta_3^8 \theta_4^8=\frac{1}{q} +232+\mathcal{O}(q)\\\nonumber
\frac{E_4}{\eta^{24}}\theta_2^8 \theta_3^8=256+\mathcal{O}(\sqrt{q}),&\qquad \qquad\frac{E_4}{\eta^{24}}\theta_2^8 \theta_4^8=256+\mathcal{O}(\sqrt{q}),\\\nonumber
\frac{E_2^2}{\eta^{24}}\sum_{a=2}^4 \theta_a(0|\tau)^{16} =\frac{2}{q}+912+\mathcal{O}(q),&\qquad \qquad\frac{E_2^2}{\eta^{24}}\theta_3^8 \theta_4^8=\frac{1}{q} -56+\mathcal{O}(q)\\\nonumber
\frac{E_4}{\eta^{24}}\theta_2^8 \theta_3^8=256+\mathcal{O}(\sqrt{q}),&\qquad \qquad\frac{E_4}{\eta^{24}}\theta_2^8 \theta_4^8=256+\mathcal{O}(\sqrt{q}),\\\nonumber
\frac{1}{\eta^{24}}\sum_{a=2}^4 \theta_a(0|\tau)^{16}(\hat{E}_2 +f_{a-1})^2=1152+\mathcal{O}(q),&\qquad \qquad\frac{\theta_3^8 \theta_4^8}{\eta^{24}} \left( (\hat{E}_2 +f_2)^2+(\hat{E}_2+f_3)^2\right)=1152+\mathcal{O}(q),\\\nonumber
\frac{\theta_2^8 \theta_4^8}{\eta^{24}} \left( (\hat{E}_2 +f_1)^2+(\hat{E}_2+f_3)^2\right)=2304+\mathcal{O}(\sqrt{q}),&\qquad \qquad\frac{\theta_2^8 \theta_3^8}{\eta^{24}} \left( (\hat{E}_2 +f_1)^2+(\hat{E}_2+f_2)^2\right)=2304+\mathcal{O}(\sqrt{q}),\\\nonumber
\frac{E_2}{\eta^{24}}\sum_{a=2}^4 \theta_a(0|\tau)^{16}({E}_2 +f_{a-1})=1440+\mathcal{O}(q),&\qquad \qquad\frac{E_2}{\eta^{24}}\theta_3^8 \theta_4^8 (2\hat{E}_2 +f_2+f_3)=-96+\mathcal{O}(q),\\\nonumber
\frac{E_2}{\eta^{24}}\theta_2^8 \theta_3^8 (2\hat{E}_2 +f_1+f_2)=\mathcal{O}(\sqrt{q}),&\qquad \qquad\frac{E_2}{\eta^{24}}\theta_2^8 \theta_4^8 (2\hat{E}_2 +f_1+f_3)=\mathcal{O}(\sqrt{q}).
\end{align}
\normalsize
%\end{eqnarray}
Finally we enlist the constant parts of the images of the different modular functions in the elliptic genus under suitable Hecke operators
\small
\begin{align}
\label{hecke1}
 \Heckea\left[\frac{\theta_3^8 \theta_4^8 E_4}{\eta^{24}} \right] &=360,\\\nonumber
 \Heckea\left[\frac{E_2^2}{\eta^{24}}\theta_3^8 \theta_4^8 \right]&=72\\\nonumber
 \Heckea \left[\frac{E_2}{\eta^{24}}\theta_3^8 \theta_4^8 (2\hat{E}_2 +f_2+f_3) \right] &=288,\\\nonumber
 \Heckea\left[\frac{\theta_3^8 \theta_4^8}{\eta^{24}} \left( (\hat{E}_2 +f_2)(\hat{E}_2+f_3)\right) \right] &=-576\\\nonumber
 \Heckea\left[\frac{\theta_3^8 \theta_4^8}{\eta^{24}} \left( (\hat{E}_2 +f_2)^2+(\hat{E}_2+f_3)^2\right) \right] &=2304
\end{align}

\begin{align}
\label{hecke2}
 \Heckeb\left[\frac{\theta_2^8 \theta_3^8 E_4}{\eta^{24}} \right] &=384,\\\nonumber
 \Heckeb\left[\frac{E_2^2}{\eta^{24}}\theta_2^8 \theta_3^8 \right]&=384,\\\nonumber
 \Heckeb \left[\frac{E_2}{\eta^{24}}\theta_2^8 \theta_3^8 (2\hat{E}_2 +f_1+f_2) \right] &=1152,\\\nonumber
 \Heckeb\left[\frac{\theta_2^8 \theta_3^8}{\eta^{24}} \left( (\hat{E}_2 +f_1)(\hat{E}_2+f_2)\right) \right] &=3456\\\nonumber
\end{align}

\begin{align}
\label{hecke3}
 \Heckec\left[\frac{\theta_2^8 \theta_4^8 E_4}{\eta^{24}} \right] &=384,\\\nonumber
 \Heckec\left[\frac{E_2^2}{\eta^{24}}\theta_2^8 \theta_4^8 \right]&=384,\\\nonumber
 \Heckec \left[\frac{E_2}{\eta^{24}}\theta_2^8 \theta_4^8 (2\hat{E}_2 +f_1+f_3) \right] &=1152,\\\nonumber
 \Heckec\left[\frac{\theta_2^8 \theta_4^8}{\eta^{24}} \left( (\hat{E}_2 +f_1)(\hat{E}_2+f_3)\right) \right] &=3456\\\nonumber
\end{align}
\normalsize

%=========================================================================

\end{document}